\documentclass[12pt]{article}

\usepackage{epsfig}
\usepackage{cite}
\usepackage{amsmath, amssymb, amsfonts}
\usepackage{color}
\usepackage{latexsym}
\usepackage{graphicx}
\usepackage[colorlinks,bookmarks]{hyperref}
\hypersetup{pdfpagemode=UseNone, pdfstartview=FitH, linkcolor=blue,
            citecolor=red, urlcolor=blue}

\def\be{\begin{equation}}
\def\ee{\end{equation}}
\def\ba{\begin{eqnarray}}
\def\ea{\end{eqnarray}}

\def\bdm{\begin{displaymath}}
\def\edm{\end{displaymath}}
\def\la{~\mbox{\raisebox{-.6ex}{$\stackrel{<}{\sim}$}}~}
\def\ga{~\mbox{\raisebox{-.6ex}{$\stackrel{>}{\sim}$}}~}
\def\bq{\begin{quote}}
\def\eq{\end{quote}}

 at 10truept
 at 14truept

\newcommand{\p}{\partial}

\newcommand{\Mpl}{M_{\mathrm{Pl}}}
\newcommand{\mps}{M_{\mathrm{Pl}}^2}

\newcommand{\bea}{\begin{eqnarray}}
\newcommand{\eea}{\end{eqnarray}}

\newcommand{\bi}{\begin{itemize}}
\newcommand{\ei}{\end{itemize}}

\newcommand{\beq}{\begin{equation}}
\newcommand{\eeq}{\end{equation}}
\newcommand{\beqa}{\begin{eqnarray}}
\newcommand{\eeqa}{\end{eqnarray}}
\newcommand{\mpl}{\Mpl}

\def\la{~\mbox{\raisebox{-.6ex}{$\stackrel{<}{\sim}$}}~}
\def\ga{~\mbox{\raisebox{-.6ex}{$\stackrel{>}{\sim}$}}~}

\def\ltap{\ \raise.3ex\hbox{$<$\kern-.75em\lower1ex\hbox{$\sim$}}\ }
\def\gtap{\ \raise.3ex\hbox{$>$\kern-.75em\lower1ex\hbox{$\sim$}}\ }
\def\gl{\ \raise.5ex\hbox{$>$}\kern-.8em\lower.5ex\hbox{$<$}\ }
\def\roughly#1{\raise.3ex\hbox{$#1$\kern-.75em\lower1ex\hbox{$\sim$}}}

\begin{document}

\thispagestyle{empty}
\begin{flushright}
September 2023 \\
DESY-23-123
\end{flushright}
\vspace*{1.25cm}
\begin{center}

{\Large \bf Implications of Weak Gravity Conjecture for}
\vskip.3cm
{\Large \bf de Sitter  Decay by Flux Discharge}  

\vspace*{1cm} {\large 
Nemanja Kaloper$^{a, }$\footnote{\tt
kaloper@physics.ucdavis.edu} and 
Alexander Westphal$^{b,}$\footnote{\tt alexander.westphal@desy.de}}\\
\vspace{.2cm} 
{\em $^a$QMAP, Department of Physics and Astronomy, University of
California}\\
\vspace{.05cm}{\em Davis, CA 95616, USA}\\
\vspace{.2cm} $^b${\em Deutsches Elektronen-Synchrotron DESY, Notkestr. 85, 22607 Hamburg, Germany}\\

\vspace{1.5cm} ABSTRACT
\end{center}
We examine implications of the weak gravity conjecture for the mechanisms for discharging cosmological
constant via membrane nucleations. Once screening fluxes and membranes which source them enter, 
and weak gravity bounds are enforced, a generic de Sitter space \underline{must} be unstable. 
We show that when all the flux terms which screen and discharge the cosmological 
constant are dominated by quadratic and higher order terms, the bounds from weak gravity conjecture 
and naturalness lead toward anthropic outcomes. In contrast, when the flux sectors are dominated 
by linear flux terms, anthropics may be avoided, and the cosmological constant may naturally 
decay toward smallest possible values. 

\vfill \setcounter{page}{0} \setcounter{footnote}{0}

\vspace{1cm}
\newpage

\section{Introduction}

In the recent sequence of papers \cite{Kaloper:2022oqv,Kaloper:2022utc,Kaloper:2022jpv,Kaloper:2023xfl,Kaloper:2023kua}
we have initiated a novel program for addressing the cosmological constant problem and de Sitter space decay by using 
discharge of $4$-form fluxes which screen, and then relax the total cosmological constant. As a result de Sitter space in this
approach is intrinsically unstable. It decays into a fractal-like patchwork of regions of ever smaller constant curvature. 
In its simplest incarnation, our approach is a generalization of the idea pioneered in 
\cite{Brown:1987dd,Brown:1988kg} about discharging cosmological constant by nonperturbative membrane nucleation processes, 
and also benefits from results of \cite{Baum:1983iwr,Hawking:1984hk,Duncan:1989ug,Bousso:2000xa,Feng:2000if}. 

Let us briefly summarize the main features of the `mechanics' of flux discharge here. Regardless of the details
of the flux sector, when the cosmological constant is very large, larger than a certain scale $\Lambda_*$ set by the 
membrane quantum numbers, the discharge proceeds by nucleation of membranes
whose size is comparable to the environment curvature radius - i.e. the horizon size. During this
``boiling" stage the discharge is essentially unsuppressed. One could try a quick estimate 
$S_{\tt bounce} \simeq - \frac{12\pi^2 \mpl^4 \Delta \Lambda}{\Lambda_{out} \Lambda_{in}}$, which is 
${\cal O}(1)$ at the cutoff \cite{Kaloper:2022oqv,Kaloper:2022utc,Kaloper:2022jpv}, and which resembles 
the Hawking-Moss instanton action \cite{Hawking:1981fz}, which supports that 
nucleation rates are not suppressed. 
A more incisive 
analysis of $S_{\tt bounce}$ actually shows that $\Lambda \rightarrow \infty$ and $\Lambda = 0$ are branch points of $S_{\tt bounce}$ as opposed to poles, and so the limits are more delicate. We will show here 
that when $\Lambda \rightarrow \infty$, $S_{\tt bounce} \rightarrow 0$, %
and so in this regime the decay rate is 
\be
\Gamma \rightarrow 1 \, ,
\label{hawkmoss}
\ee
which indicates barrier-less tunneling in the Euclidean theory. This could happen 
in the limits of the well known analyses of \cite{Coleman:1980aw,Parke:1982pm}, for very fast 
bubble nucleations \cite{Guth:1982pn,Turner:1992tz,Freese:2004vs}. In the thin wall limit
of tunneling between a false and a true vacuum for a scalar field, the wall tension measures the
barrier area, which controls the tunneling rate. Hence when the cosmological constant dominates
over the tension terms, the barrier is negligible, which yields an unsuppressed 
nucleation rate.

Even more accurately, the rapid decay during this stage is moderated by a small bounce action, which 
eventually gradually increases toward 
$S_{\tt bounce} \simeq + \frac{24\pi^2 \mpl^4}{\Lambda_{out}}$ as 
the initial cosmological constant decreases. The resulting decay rate 
disfavors the largest possible values of the cosmological constant, and favors the smallest ones, 
as the terminal outcome, because the more curved backgrounds are
more unstable. To be relevant, this regime must involve the cosmological 
constant values below the cutoff; otherwise it is excluded from the effective theory description. When 
this holds, the cosmological constant will discharge to the smallest values achievable,
for as long as the rate remains nonzero. 
Due to the increase of the bounce action with the decrease of $\Lambda$, 
the resulting distribution of cosmological constant 
values is skewed toward the smallest possible values, that can be reached given a set of membrane charges. 
This stage is important for populating the de Sitter landscape in ways which 
reach the smallest values of the terminal
cosmological constant. 

Once $\Lambda$ decreases to below the critical scale, discharge rate dramatically slows down. 
There are important quantitative differences between the dynamics depending on whether discharge processes are 
controlled by linear $4$-form flux terms, or by  quadratic and higher power ones. 
If linear flux terms dominate discharge processes, as in \cite{Kaloper:2022oqv,Kaloper:2022utc,Kaloper:2022jpv,Kaloper:2023xfl,Kaloper:2023kua}, 
the discharge channels shut off faster, since the rates strongly depend on the initial cosmological constant.
As noted above, for these processes 
the decay rate asymptotes toward an essential singularity as $\Lambda \rightarrow 0^+$, 
\be
\Gamma \rightarrow \exp\Bigl(-\frac{24 \pi^2 \mpl^4}{\Lambda_{out}}\Bigr) \, ,
\label{atrrate}
\ee
where $\mpl^2 = 1/8\pi G_N$. This ``braking" stage protects the tiniest values of cosmological constant by
making those geometries most stable. Together with the faster discharge, ``boiling" regime preceding it, the overall 
dynamics favors de Sitter spaces with the smallest attainable cosmological constant 
\cite{Kaloper:2022oqv,Kaloper:2022utc,Kaloper:2022jpv,Kaloper:2023xfl,Kaloper:2023kua}. 
This occurs for all generic, natural values 
of the initial cosmological constant near and below the cutoff of the theory, at a scale ${\cal M}_{\tt UV} \le \mpl$. 

This is in contrast to previous works \cite{Brown:1987dd,Brown:1988kg,Hawking:1984hk,Duncan:1989ug,Bousso:2000xa,Feng:2000if} 
where the flux sector \underline{does not} include linear
$4$-form terms but starts with $F^2$. Then, whenever the initial cosmological constant is not fine tuned, but starts 
near the cutoff value $\Lambda \sim {\cal M}_{\tt UV}^4 \la \mpl^4$, the terms which control the transitions 
select decay channels which have a broad regime with rates without accumulation points, which asymptote to a constant value that depends 
on the {\it difference} of the cosmological constants in the parent and descendant bubbles, and not the individual values, so
\be
\Gamma \simeq \exp\Bigl(-\frac{27\pi^2}{2} \frac{{\cal T}^4}{(\Delta \Lambda)^3}\Bigr) 
\rightarrow \exp\Bigl(-\frac{27\pi^2}{2}  \frac{{\cal T}^4}{(2 \Lambda_{\tt QFT})^{3/2} {\cal Q}^3}\Bigr) \, .
\label{constrate}
\ee
Here $\Lambda_{\tt QFT}$ is the cosmological constant from the field theory sector which is being neutralized, 
and ${\cal T}$ and ${\cal Q}$ membrane tension and charge, respectively. 
In such cases,
one finds that the terminal distribution of cosmological constant values can be uniform if the phase space of possible
values were equiprobable, and if the preceding ``boiling" stage was not long (or completely excluded). In this case 
one can select the final value by resorting to the anthropic principle. 

The `boundary' between the ``boiling" and ``braking" stages is controlled by the ratio of the cosmological 
constant before a transition and, in general, the tension and charge of the membrane, in the units of Planck scale. 
The precise value of the critical value $\Lambda_*$ where the transition happens is detail-dependent (and actually 
can be quite broad) and we will outline the possibilities below. We stress this presupposes that both the
``boiling" and ``braking" stages are below the cutoff, within the realm of effective theory.  
This is not automatic for the ``boiling" stage. 

This argument indicates that together, naturalness in the QFT sense and any constraints on 
membrane charges and tensions relative to the cutoff 
impose restrictions on the cosmological constant cancellation via screening and membrane discharge. 
Staying below the cutoff is not only an issue of reliability
but of caution born of necessity: above the cutoff lurks the wormhole regime, which is notoriously unreliable 
\cite{Coleman:1988tj,Fischler:1988ia,Klebanov:1988eh,Coleman:1989ky}. 
So the ``brute force" tuning and retuning of parameters in the equations which arise 
in the semiclassical limit may not be arbitrarily done to evade those restrictions. In this paper, we explore these
restrictions and the model-building requirements they impose. Specifically, 
we deploy the bounds on the membrane dynamics
which arise from the Weak Gravity Conjecture (WGC) \cite{Arkani-Hamed:2006emk} and apply them to the generic
natural boroughs of the landscape, where the field theory contribution to the vacuum energy is technically natural,
of order $\Lambda_{\tt QFT} \la {\cal M}_{\tt UV}^4$. 

It should be immediately clear that the WGC bounds affect the nature of de Sitter space fundamentally 
once we introduce the flux screening and discharge mechanisms. Once the cosmological constant 
receives the contributions from fluxes, and charged tensional membranes are included, so that fluxes can
change by membrane nucleation, the WGC bounds immediately 
imply that there is no absolutely stable de Sitter space!
The reason is simple: to stop the quantum mechanically induced discharges, membranes must be decoupled, 
which means, the tension of \underline{all the membranes} that can change the cosmological constant 
must go to, formally, infinity -- or in practice, above the cutoff. This is precisely the limit prohibited by WGC 
\cite{Arkani-Hamed:2006emk}. Even if we do not violate WGC, this limit indicates that the WGC bounds,
which constrain the ratios of charges, tensions and the cutoff, will affect the specific details of discharge dynamics. 

We find that when the $4$-form sector is dominated by 
quadratic and higher powers of fluxes, the natural discharges by membranes which obey 
WGC are mediated by the same
types of instantons as in \cite{Brown:1987dd,Brown:1988kg}, which admit a broad regime 
with the asymptotic decay 
rate given by (\ref{constrate}). The  ``boiling" regime is pushed out of the range of the effective theory and 
the initial vacuum energy to be cancelled, near the QFT cutoff, is already at the edge of the ``braking" regime. Such 
setups can therefore be naturally used to provide a framework for anthropic selection of the 
terminal cosmological constant, as in \cite{Bousso:2000xa}.

If we chose to violate WGC for {\it all} charged membranes in the theory, 
the low scale attractor (\ref{atrrate}) will appear, and the ``boiling" stage may reappear below the cutoff. 
However the WGC violations required to make the setup natural are quite severe, because the charges of 
affected membranes become very small, and restoring WGC by adding membranes which 
satisfy the bounds can also restore channels for dominantly uniform
discharge, that bring back anthropics, or force more fine tuning. 

In contrast, when the linear flux terms are present, and when they dominate over higher powers in the effective
action, the dynamics changes significantly  \cite{Kaloper:2022oqv,Kaloper:2022utc,Kaloper:2022jpv,Kaloper:2023xfl,Kaloper:2023kua}.
When the fluxes are natural, of the order of cutoff-scale QFT 
vacuum energy $\Lambda_{\tt QFT} \simeq {\cal M}_{\tt UV}^4$, and they 
satisfy WGC, both the ``boiling" and ``braking"
stages are below the cutoff, and they, together, favor 
the nucleation of a sequence of bubbles ending with the smallest possible
terminal cosmological constant. Hence in this case the universes with tiny cosmological constant could  
arise naturally, 
without invoking anthropic reasoning. We also note, that even if some 
of the membranes involved are discharged by processes dominated by quadratic flux 
terms, the outcome may  remain 
unaffected as long as there are many decay channels dominated by linear flux terms. A more detailed investigation
of the evolution which involves a variety of decay channels, of both types mentioned here and with a range
of charges and tensions is therefore warranted. 

The paper is organized as follows: in the next section we review WGC, starting with point particles and then provide a general set of inequalities for
charged membranes in four dimensions. Next, in Section 3. we discuss the effective action for fluxes and charged tensional membranes 
using magnetic duals of $4$-forms, and revisit the mechanics of flux discharges of  \cite{Kaloper:2022oqv,Kaloper:2022utc,Kaloper:2022jpv,Kaloper:2023xfl,Kaloper:2023kua}. In Section 4. we turn to the implications
of WGC and naturalness for flux discharge processes and explain the limitations which arise for effective theory description of de Sitter
decay. We give a summary of the results and discuss open questions in the last section. 

\section{WGC in a Nutshell: a Lightning Review} 

If objects supporting event horizons were really forever, the retrieval of information 
about the material that went into them may be impossible. In quantum physics, this suggests 
that event horizons may catalyze unitarity loss, and hence endanger and obstruct 
quantum mechanics itself \cite{Hawking:1976ra}. Preempting this implies subtle consistency conditions on models 
of matter coupled to quantum gravity. A specific application concerns charged black holes. 
As is well known, generic black holes are actually not black since they radiate 
like black bodies at Hawking temperature. To ensure that they radiate out the charge that 
went in with the material which formed a black hole, it is necessary that 
there are sufficiently light charged particles that can stream outside. 
This imposes a condition on charge per unit mass \cite{Arkani-Hamed:2006emk}, which is now called the electric WGC: 
for each conserved gauge charge there must be a sufficiently light charge carrier such that 
\be
\frac{e}{m} \ge  \sqrt{G_N} \, ,
\label{wgc} 
\ee
where $e$ and $m$ are the carrier charge and mass. 
This can be deduced very simply from conservation laws \cite{Banks:2006mm,Palti:2019pca}: 
consider a black hole of mass $M$ with charge $Q$, where by conservation
of mass and charge $M \ge \sum_i m_i$ (as we allow for energy contribution from neutral sources) and $Q = \sum_i e_i$. Thus
\be
\frac{M}{Q} = \frac{1}{Q} \sum m_i = \frac{1}{Q} \sum \frac{m_i}{e_i} e_i 
\ge \frac{1}{Q} \bigl(\frac{m}{e}\bigr)_{min} \sum e_i = \bigl(\frac{m}{e}\bigr)_{min} \, .
\label{gbound}
\ee
Applying this to extremal black holes $M = Q/\sqrt{G_N}$, which have the largest ratio $Q/M$ due to the horizon regularity constraints
yields the strongest bound: Eq. (\ref{wgc}), precisely. 

However many charged black holes can become ultra-cold in the extremal limit, and cease to emit 
Hawking radiation. If they were to 
last forever, they would cause problems behaving as troublesome remnants \cite{Susskind:1995da}. 
Yet even if Hawking radiation ceases there are nonperturbative, non-thermal processes which 
lend to black hole discharge \cite{Gibbons:1975kk}. Essentially, these are variants of Schwinger charged 
particle production in background electric fields \cite{Schwinger:1951nm}. 
This decay channel arises thanks to Heisenberg uncertainty principle, whereby a particle-antiparticle pair emerges in an external electric field $E$. The field accelerates virtual particles in the pair away from each other and transfers enough energy to them
that they can get on shell instead of annihilating away. 

A very nice intuitive argument is given in \cite{Khriplovich:1999gm}, building on the work of \cite{Sauter:1931zz}, 
which we revisit here. We will model the pair creation and their separation due to the work of the background field 
as the process of initially accelerating a negative energy ``virtual" particle, which gains enough energy due to the acceleration to 
become a positive energy particle that propagates away, leaving behind a ``hole" --  a positive energy antiparticle after
charge conjugation -- that propagates away in the opposite direction. Working in the rest frame of one of the pair, 
which is also initially the rest frame of the pair, the dispersion relation 
after a small displacement $\delta z$ is $(\varepsilon + eE\delta z)^2 - \vec p^2 = m^2$. Solving for $p_z$, with the initial condition
$\varepsilon = - m$ at $\delta z = 0$ (recall that $c = \hbar = k_B = 1$), 
\be
p_z = \sqrt{(-m + eE\delta z)^2 - m^2} = \sqrt{eE\delta z(eE\delta z - 2m)} \, .
\ee

Clearly, the square root vanishes at $\delta z = 0$ and $\delta z = 2m/eE$. In between these two locations, the square root 
is imaginary, and so the Euclidean momentum $\pi_z = -i p_z$ is real, $\pi_z =  \sqrt{eE\delta z(2m- eE\delta z)}$. In this regime the
particle is ``virtual", with imaginary momentum, being accelerated by the electric  field $E$ toward positive energies. Since the mass shells are tilted by the
electric field potential energy, the particle can tunnel from the negative mass shell to the positive one, 
and subsequently propagate out to infinity \cite{Khriplovich:1999gm}. 
This occurs when the Euclidean momentum vanishes at $\delta z = 2m/eE$, which can be understood 
as the instant where the particle gains enough energy through the work of the electric field. Indeed, 
since $\delta W \simeq e E \delta z$, $\delta W \simeq 2 m$ implies that enough energy is transferred at $\delta z \simeq 2m/eE$. 

We can estimate the particle production rate to the leading order by employing WKB approximation, and computing the 
Euclidean action by integrating over the region where $p_z$ is imaginary (the ``barrier"):
\be
S_{\tt E} = \int_0^{2m/eE} dz \, \pi_z = \frac{\pi}{2} \, \frac{m^2}{eE} \, , 
\ee
to get the tunneling wavefunction $\Psi = e^{-S_{\tt E}}$. The rate is given by $\Gamma \sim |\Psi|^2$, hence
\be
\Gamma \simeq e^{-\pi m^2/eE} \, .
\ee
In weak electric fields $E \rightarrow 0$, the rate goes to zero, while for strong fields $E > \pi m^2/e$ the exponential suppression
disappears, and the rate is polynomially fast. 

Applying this formula to a charged black hole, and taking the strongest electric field available just 
outside of the outer 
event horizon, yields $E = Q/r_+^2$ where $Q$ is the black hole charge and $r_+ = G_N M + \sqrt{G_N^2 M^2 - G_N Q^2}$, where $M$ is the black hole mass, $Q$ its charge, and $G_N$ Newton's constant. Therefore 
\be
\Gamma \simeq e^{-\pi m^2r_+^2 /eQ} \, .
\ee
It turns out that, despite technical subtleties, this equation gives the correct leading order decay rate describing black hole
discharge due to nonperturbative quantum effects \cite{Gibbons:1975kk,Khriplovich:1999gm}. Note, that these processes do not
cease in the extremal limit, and that discharge continues even when $M = Q/\sqrt{G_N}$. Further note, that while these processes 
are slow for large black holes, they speed up as the mass decreases. They can also be augmented by spurts of Hawking radiation that can restart
the charge loss by light particle emission, and
go faster when charge carriers are light, $m \ll \mpl$. But at least in principle, as long as the Euclidean 
action $S_{\tt E}$ can continuously decrease to $\le 1$, the discharge can proceed -- and speed up near the end -- with the
black hole disappearing. As 
shown in \cite{Banks:2006mm}, reaching $m^2r_+^2 /eQ < 1$ is inevitable as long as Eq. (\ref{wgc}) holds. To see this, 
substitute $r_+ = G_N M + \sqrt{G_N^2 M^2 - G_N Q^2}$ into $m^2r_+^2 /eQ < 1$. 
The resulting inequality after squaring it up and manipulating terms as in \cite{Banks:2006mm} becomes
\be
Q^2 + \frac{eQ}{G_N m^2} \ge 2\frac{M}{m} \sqrt{eQ} \, .
\label{qmeq}
\ee
Next, demand $M \ge Q m/e$ to ensure that kinematical constraints
can be met, for simplicity define $Q = \zeta^2$ and divide everything by $\zeta^2$ \cite{Banks:2006mm}; this maps (\ref{qmeq}) to 
\be
f(\zeta) \ge 0 \, , ~~~~~~~~~~~ f(\zeta) = 1 + \frac{e}{G_N m^2  \zeta^2} - \frac{2}{\sqrt{e}{\zeta}} \, .
\label{qmeqs}
\ee
The function $f(\zeta)$ has two zeros, one at $\zeta = 0$ and the other away, at a location approximately determined by
$\zeta_0 \simeq e^{3/2}/2 G_N m^2$. In between these two values $f$ is negative, and so the inequality cannot be satisfied there. To satisfy the
inequality, $\zeta$ must exceed  $\zeta_0$. However, as the black hole charge $\zeta = Q^2$ decreases from some large initial value, the root $\zeta_0$ 
must approach the origin, which means that at fixed $e$, the mass 
$m$ must be dialed up to satisfy $f \ge 0$ -- eventually running afoul of Eq. (\ref{wgc}). The fastest way 
for this to occur is along the parameter space curve extremizing $f$ in the $\zeta$ direction, which implies $\zeta_{\tt max} = e^{3/2} G_N m^2$.
Hence the strongest bound comes from imposing (\ref{qmeq}) at this value of $\zeta$. Substituting in (\ref{qmeq}), we find that
$f \ge 0$ implies
\be
f(\zeta_{\tt max}) = 1 - \frac{G_N m^2}{e^2} \ge 0  ~~~~~~ \implies ~~~~~~ \frac{\sqrt{G_N} m}{e} \le 1 \, , 
\ee
which is precisely the same as the bound of Eqs. (\ref{wgc}), (\ref{gbound}). 
This implies that as long as there are light particles carrying charge $e$ which obey (\ref{wgc}), 
charged black holes cannot linger forever. Both perturbative and nonperturbative particle production processes can
discharge them. Conversely, if (\ref{wgc}) does not hold for any charged species, neither discharge channel will be generally accessible,
and remnants, and perhaps other problems, would seem to be difficult to avoid \cite{Arkani-Hamed:2006emk,Banks:2006mm}.
Eq. (\ref{wgc}) provides protection from such problems. 

There is also a magnetic variant of WGC, which deals with the interplay of magnetic solitons with gravity \cite{Arkani-Hamed:2006emk}
(see also \cite{Hebecker:2015zss}). An issue here is that in the weak coupling regime of gauge theory magnetic monopoles are very heavy,
with the mass $m_{\tt monopole} \sim {\cal M}_{\tt UV}/e^2$, where ${\cal M}_{\tt UV}$ is the UV cutoff of the theory. Combining this with
the WGC bound $m_{\tt monopole} \le e_{\tt magnetic} \sqrt{G_N} \simeq \sqrt{G_N}/e$ yields 
\be
{\cal M}_{\tt UV} \le e \sqrt{G_N} \, ,
\ee
which means that the cutoff of a weakly-coupled gauge theory must be below the Planck scale. This ensures that the monopole
is not a black hole: combining $m_{\tt monopole} \sim {\cal M}_{\tt UV}/e^2$ with the size of the monopole core 
$R_{\tt monopole} \sim 1/{\cal M}_{\tt UV}$ yields immediately $m_{\tt monopole} \le G_N R_{\tt monopole}$ \cite{Arkani-Hamed:2006emk}, violating the hoop conjecture which black holes satisfy. 

The WGC bounds discussed here for point particles can be generalized for extended objects. Specifically, we will be interested
in the implications of charged tensional membranes in four dimensions. For them, the electric weak gravity bound 
generalizes to
\be
\frac{{\cal Q}}{{\cal T}} \ge {\sqrt{G_N}} \, .
\label{wgcmeme}
\ee
The statement of WGC then is that in the spectrum of the theory which includes charged membranes, for each type 
of charge there must be at least one membrane which satisfies the inequality (\ref{wgcmeme}). The magnetic form of the bound 
is a bit more subtle, having been deduced \cite{Hebecker:2015zss} to be the bound on the cutoff of the theory
\be
{\cal M}_{\tt UV} \le \frac{{\cal Q}^{1/3}}{G_N^{1/6}} \, ,
\label{wgcmemm}
\ee
found by estimating the membrane tension in more than $4D$ by the energy stored in the field sourced by ${\cal Q}$, using the expected
scaling of the gravitational radius with the higher-dimensional gravitational constant, requiring that there exists 
a magnetic membrane without the horizon, and then dimensionally reducing the result to $4D$. 
The inequalities (\ref{wgcmeme}), (\ref{wgcmemm}) will play important role in our
arguments in what follows. 

\section{Discharges with Linear and Quadratic Flux Terms}

General theories of $4$-forms coupled to gravity and sourced by charged tensional membranes were examined in 
 \cite{Kaloper:2023xfl,Kaloper:2023kua}. They split into two qualitatively 
 different classes, depending on whether the linear 
 $4$-form flux terms are present and dominant in the action, or not. 
 For this reason we will focus here on the more special
 limiting forms, comprised of only linear and quadratic terms, which simplifies the discussion without any loss of generality. 
Further, the technical analysis simplifies by replacing the $4$-form with its magnetic dual, ${\cal F} \leftrightarrow  {{^\ast}\lambda}$, 
and replacing the $4$-form Lagrangian ${\cal L}({\cal F})$ with its Legendre transform ${\cal L}(\lambda)$  \cite{Dvali:2005an,Kaloper:2020jso}. 
As explained in \cite{Kaloper:2023xfl}, this amounts to the Routhian transformation of the theory.
Concretely, we start with 
\be
S = \int d^4x \Bigl\{ \sqrt{g} \Bigl(\frac{\mpl^2}{2} R  - {\cal L}_{\tt QFT} - \frac{1}{48} {\cal F}_{\mu\nu\lambda\sigma}^2 \Bigr)
- \frac{\alpha}{24} \epsilon^{\mu\nu\lambda\sigma} {\cal F}_{\mu\nu\lambda\sigma} \Bigr\} \, ,
 \label{linsquare} 
\ee
motivated by, e.g. \cite{Aurilia:1980xj}, where 
$\alpha$ is a fixed $4$-form theory coupling parameter induced by nontrivial CP-breaking effects
\cite{Aurilia:1980xj}. We then add 
the boundary term $\int d^4x \frac{1}{3} \epsilon^{\mu\nu\lambda\sigma} \partial_{\mu} (\lambda \, {\cal A}_{\nu\lambda\sigma})$ 
to (\ref{linsquare}), define the new variable $\tilde {\cal F}_{\mu\nu\lambda\sigma} = {\cal F}_{\mu\nu\lambda\sigma} - (2 \lambda - \alpha) {\sqrt{g}} {\epsilon_{\mu\nu\lambda\sigma}} $ 
and integrate $\tilde{\cal F}$ out. The resulting ``bulk" action is \cite{Kaloper:2023xfl}
\be
S = \int d^4x \Bigl\{\sqrt{g} \Bigl(\frac{\mpl^2}{2} R  - {\cal L}_{\tt QFT} - {2} \bigl(\lambda-\frac{\alpha}{2})^2 \Bigr) 
+ \frac{1}{3} \epsilon^{\mu\nu\lambda\sigma} \partial_\mu \Bigl(\lambda\Bigr)  {\cal A}_{\nu\lambda\sigma} \Bigr\}  \, .
\label{linsquare3}  
\ee
We next expand ${2} \bigl(\lambda-\frac{\alpha}{2})^2 = 2\lambda^2 - 2 \alpha \lambda + \alpha^2/2$, and absorb the
flux-independent term $\alpha^2/2$ into the QFT vacuum energy, ${\cal L}_{\tt QFT} + \alpha^2/2 \rightarrow {\cal L}_{\tt QFT}$. 
Further, since $4$-form should be viewed as a higher rank gauge theory, we add to (\ref{linsquare3}) 
the gauge field charges -- the charged tensional membranes -- and boundary terms required to properly provide junction
conditions across the membrane walls. This is motivated by the general lore that quantum gravity does not 
coexist with global symmetries \cite{Banks:1991mb,Banks:2010zn}, and without charges the $4$-form theory would in fact admit generalized higher-form symmetries. When charges are present, those are broken by gauge currents \cite{Gaiotto:2014kfa,Reece:2023czb}. 

Finally we parameterize $2\alpha = c_1 {\cal M}_{\tt UV}^2$, and note that the value of $c_1$ controls how close this  term is to the cutoff. Such terms arise naturally in axion physics, when the 
CP-violating effects in some non-trivial gauge theory
max out \cite{Aurilia:1980xj}. We could even imagine that such a theory has an axion with a very large decay constant 
$f \ga \mpl$ and where quantum gravity effects break shift symmetry very strongly. In any case, the final effective action 
for a single gauge sector coupled to gravity, which we
will use in what follows, is 
\ba
S &=& \int d^4x \Bigl\{\sqrt{g} \Bigl(\frac{\mpl^2}{2} R 
- {\cal L}_{\tt QFT} + c_1{\cal M}_{\tt UV}^2\lambda- 2 \lambda^2 \Bigr) + 
\frac{1}{3} {\epsilon^{\mu\nu\lambda\sigma}} \partial_\mu \Bigl(\lambda \Bigr) {\cal A}_{\nu\lambda\sigma} \Bigr\} \nonumber \\
&-&  \int d^3 \xi \sqrt{\gamma} \mpl^2[ K ] - {\cal T}_{\cal A} \int d^3 \xi \sqrt{\gamma}_{\cal A} - {\cal Q}_{\cal A} \int {\cal A}  \, ,
\label{actionnewmemdcharg} 
\ea
where ${\cal T}_{\cal A}$ and ${\cal Q}_{\cal A}$ are the membrane tension and charge, respectively, 
the term $\propto K$ is the Israel-Gibbons-Hawking term 
for gravity which encodes boundary conditions across membrane walls, and $[...]$ is the jump across a membrane. The coordinates
$\xi$ are coordinates along a membrane worldvolume, embedding it in spacetime. The charge terms are 
\be
\int {\cal A} = \frac16 \int d^3 \xi {\cal A}_{\mu\nu\lambda} \frac{\p x^\mu}{\p \xi^\alpha} \frac{\p x^\nu}{\p \xi^\beta} 
\frac{\p x^\lambda}{\p \xi^\gamma} \epsilon^{\alpha\beta\gamma} \, .
\ee
We take ${\cal T}_{\cal A} > 0$ to avoid problems with ghosts and negative energies. This is a special case of actions discussed in
\cite{Kaloper:2023xfl}, which suffices for our purposes here. 

To study the discharge processes, we Wick-rotate the action (\ref{actionnewmemdcharg}) to Euclidean time. This Euclidean action 
controls the nucleation rates $\Gamma \sim e^{-S_E}$ \cite{Coleman:1980aw}.  The analysis is given in detail 
in \cite{Kaloper:2022oqv,Kaloper:2022utc,Kaloper:2022jpv,Kaloper:2023xfl}, and we just summarize it here. To transition
to Euclidean picture, we replace $t = - i x^0_E$, which gives 
$- i \int d^4x \sqrt{g} {\cal L}_{\tt QFT} = - \int d^4x_E \sqrt{g} {\cal L}^E_{{\tt QFT}}$; with the standard conventions, 
${\cal A}_{0 jk} = {\cal A}^{E}_{0jk}$, ${\cal A}_{jkl} =  {\cal A}^{E}_{jkl}$ we get ${\cal F}_{\mu\nu\lambda\sigma} = {\cal F}^{E}_{\mu\nu\lambda\sigma}$, 
and $\epsilon_{0ijk} = \epsilon^{E}_{0ijk}$, $\epsilon^{0ijk} = -\epsilon_E^{0ijk}$. The membrane source terms transform to
$- i {\cal T}_{\cal A} \int d^3 \xi \sqrt{\gamma} = - {\cal T}_{\cal A} \int d^3 \xi_E \sqrt{\gamma}$ and $i {\cal Q}_{\cal A} \int {\cal A}_i = - {\cal Q}_{\cal A} \int {\cal A}_i$. The Euclidean action by $i S = - S_E$ is (below we drop the subscript ${E}$): 
\ba
S_E&=&\int d^4x_E \Bigl\{\sqrt{g} \Bigl(-\frac{\mpl^2}{2} R_E + {\cal L}(\lambda) 
+ \Lambda_{\tt QFT} - c_1 {\cal M}_{\tt UV}^2 \lambda + 2 \lambda^2 \Bigr) + \frac{1}{3} {\epsilon^{\mu\nu\lambda\sigma}_E} \partial_\mu \Bigl( \lambda \Bigr) {\cal A}^E_{\nu\lambda\sigma} \nonumber \\
\label{actionnewmemeu}
&+&\int d^3 \xi \sqrt{\gamma} \mpl^2[ K_E ]  + {\cal T}_{\cal A} \int d^3 \xi_E \sqrt{\gamma}_{\cal A} - \frac{{\cal Q}_{\cal A}}{6} \int d^3 \xi_E \, {\cal A}^E_{\mu\nu\lambda} \, 
\frac{\p x^\mu}{\p \xi^\alpha} \frac{\p x^\nu}{\p \xi^\beta} 
\frac{\p x^\lambda}{\p \xi^\gamma} \epsilon_E^{\alpha\beta\gamma} \, .
\ea

In the action we set $\langle {\cal L}^E_{\tt QFT} \rangle = \Lambda_{\tt QFT}$, 
with $\Lambda_{\tt QFT}$ the regulated matter sector vacuum energy to an arbitrary loop order, since 
we consider transitions on backgrounds with local $O(4)$ symmetry that have minimal Euclidean action 
and dominate the evolution \cite{Coleman:1980aw,Parke:1982pm}. 
When the QFT vacuum energy is natural, QFT/gravity couplings imply 
$\Lambda_{\tt QFT} = {\cal M}_{\tt UV}^4 
+ \ldots \equiv \mps  \lambda_{\tt QFT}$, where
${\cal M}_{\tt UV}^4$ is the QFT UV cutoff and ellipsis denote sub-leading terms \cite{Englert:1975wj,Arkani-Hamed:2000hpr}. 
Hence the total cosmological constant in any bulk patch is 
\be
\Lambda_{\tt total} = \Lambda_{\tt QFT} - c_1 {\cal M}_{\tt UV}^2 \lambda + 2 \lambda^2  \, , 
\label{totalcc}
\ee
where $\lambda$ can vary from patch to patch across membrane walls. 

A nucleation of a membrane changes the flux of $\lambda$ inside it, and hence the total cosmological constant in the interior. 
The resulting geometry comprises of two de Sitter patches glued along the membrane, with tension and charge controlling
the membrane-sourced discontinuity. Away from the membrane, de Sitter patches are described with the metrics
\be
ds^2_E =  dr^2 + a^2(r) \, d\Omega_3 \, ,
\label{metricsmax}
\ee
where $d\Omega_3$ is the line element on a unit $S^3$. The warp factor $a$ is the solution of the Euclidean ``Friedmann equation", 
\be
3 \mpl^2 \Bigl( \bigl(\frac{a'}{a}\bigr)^2 - \frac{1}{a^2} \Bigr) = - \Lambda_{\tt total} \, .
\label{fried}
\ee
The prime designates an  $r$-derivative. From here on, we will drop the subscript ``${\tt total}$". 
The boundary conditions induced on a membrane for gauge fields and gravity are \cite{Kaloper:2022oqv,Kaloper:2022utc,Kaloper:2022jpv} 
\be
a_{out} = a_{in} \, , ~~~~~~ \lambda_{out} - \lambda_{in}  = \frac{{\cal Q}_{\cal A}}{2} \, ,  ~~~~~~ \mps \Bigl(\frac{a_{out}'}{a} - \frac{a_{in}'}{a} \Bigr)
 = -\frac{{\cal T}_{\cal A}}{2} \, ,
 \label{metricjc}
\ee 
in the coordinate system where the outward membrane normal vector is oriented in the direction of the 
radial coordinate; $r$ measures the distance in this direction. Subscripts $``out"$ and $``in"$ refer to the membrane's exterior 
(``parent de Sitter") and interior (``descendant de Sitter"), respectively. 
The discontinuities in $\lambda$ and $a'$ follow since a membrane is a Dirac $\delta$-function source of charge and tension. 

We proceed by solving (\ref{fried}) for $a' = \zeta_j \sqrt{1-  \frac{\Lambda a^2}{3 \mpl^2}}$, where 
$\zeta_j = \pm 1$ designate the two possible branches of the square root. Using this and the junction conditions for the
magnetic fluxes (\ref{metricjc}), the value of $\Lambda_{\tt QFT}$ cancels out and we obtain \cite{Kaloper:2022oqv,Kaloper:2022utc,Kaloper:2022jpv,Kaloper:2023xfl,Kaloper:2023kua}
\ba
\zeta_{out} \sqrt{ 1-  \frac{\Lambda_{out} a^2}{3 \mps}} 
&=& - \frac{{\cal T}_{\cal A}}{4\mps}\Bigl(1 + \frac{2\mpl^2 {\cal M}_{\tt UV}^2 {\cal Q}_{\cal A}}{3 {\cal T}^2_{\cal A}} \bigl({c_1} - 4 \frac{\lambda}{{\cal M}_{\tt UV}^2} \bigr)\Bigr)\, a \, , \nonumber \\
\zeta_{in} \sqrt{1-  \frac{\Lambda_{in} a^2}{3\mps}} 
&=& \frac{{\cal T}_{\cal A}}{4\mps}\Bigl(1 -  \frac{2\mpl^2{\cal M}_{\tt UV}^2{\cal Q}_{\cal A}}{3 {\cal T}^2_{\cal A}} \bigl({c_1}  - 4 \frac{\lambda}{{\cal M}_{\tt UV}^2})\Bigr) \, a \, , 
\label{diffroots}
\ea
where we take the flux to be made up of a large number of charge units, $\lambda \gg {\cal Q}_{\cal A}$. If this were not so, 
we would replace $\lambda \rightarrow \lambda_{out} - {\cal Q}_{\cal A}/4$ in (\ref{diffroots}) (in the large flux case, the distinction between
the $``in"$ and $``out"$ fluxes in these equations is irrelevant). 

\begin{figure}[thb]
    \centering
    \includegraphics[width=13cm]{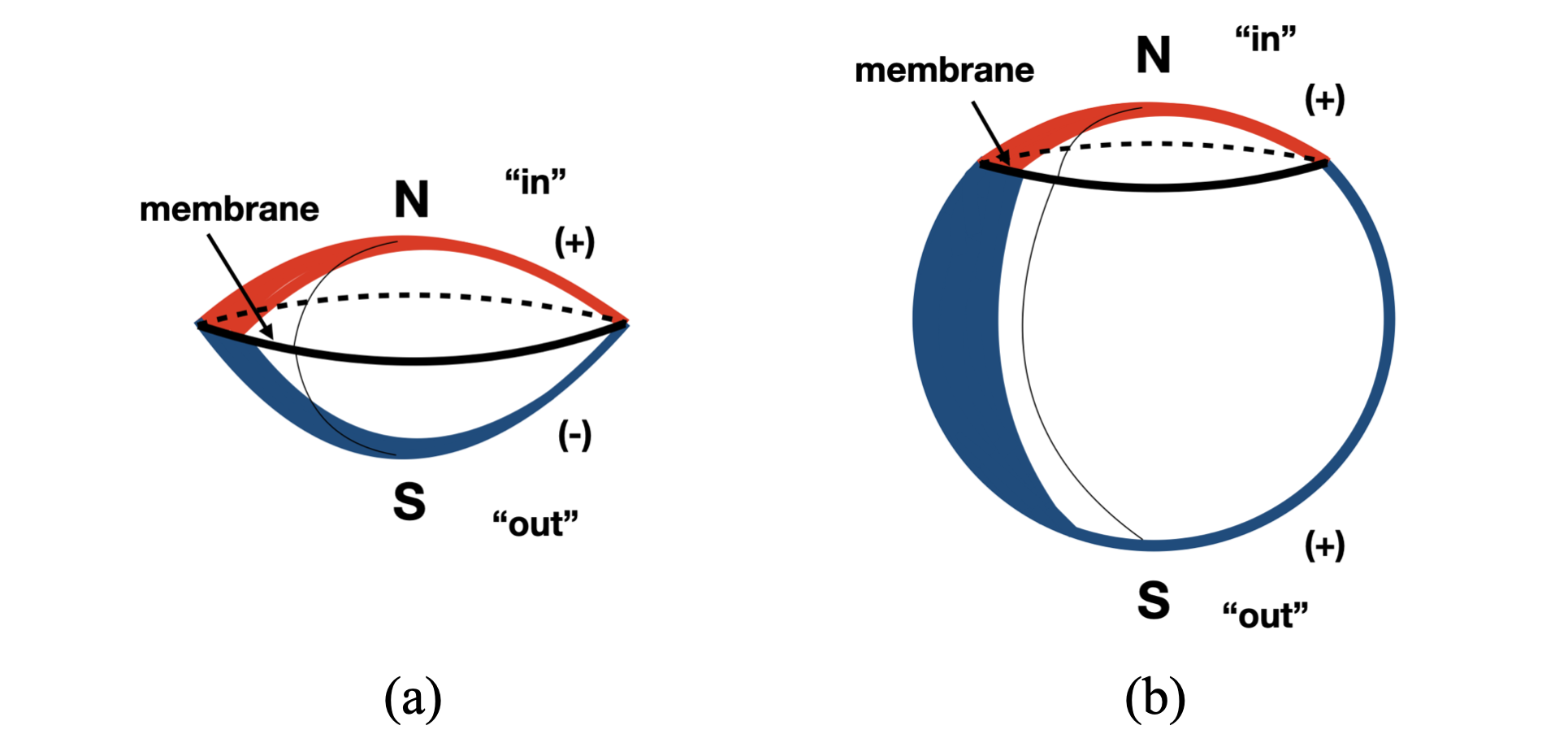} 
     \caption{(a): a $|q| <1$ instanton mediating $dS \rightarrow dS$ with $(\zeta_{out}, \zeta_{in}) = (-, +)$. 
    (b): a large flux, $|q| >1$ instanton mediating $dS \rightarrow dS$ with $(\zeta_{out}, \zeta_{in}) = (+, +)$.}
    \label{fig1}
\end{figure}

The equations (\ref{diffroots}) play a crucial role, since they select the membrane discharge channel which relaxes the vacuum energy,
and control relaxation dynamics. The point is, that the right hand side (RHS) of (\ref{diffroots}) can be written as 
$\mp\frac{{\cal T}_{\cal A}}{4\mps}\Bigl(1 \pm q \Bigr)$ where 
\be
q = \frac{2\mpl^2 {\cal M}_{\tt UV}^2 {\cal Q}_{\cal A}}{3 {\cal T}^2_{\cal A}} 
\bigl({c_1} - 4 \frac{\lambda}{{\cal M}_{\tt UV}^2} \bigr)  \, . 
\label{qeq}
\ee
Thus when $|q|<1$, the terms in the parenthesis on the RHS  of (\ref{diffroots}) keep the same sign, and Eqs. (\ref{diffroots}) only have solutions
when $(\zeta_{out}, \zeta_{in}) = (-, +)$. Conversely, when $|q|>1$, the solutions describing $dS \rightarrow dS$ transitions can only be found when 
 $(\zeta_{out}, \zeta_{in}) = (+, +)$ or $(-,-)$. Of the latter two cases,  $(\zeta_{out}, \zeta_{in}) = (+, +)$ dominates over 
 $(\zeta_{out}, \zeta_{in}) = (-, -)$  because it has smaller Euclidean action. The instantons mediating $(-,+)$ and $(+,+)$ processes are given in
 Fig. (\ref{fig1}). 

Which of the two channels is selected by the boundary conditions has a dramatic effect on the discharge dynamics. 
The bounce action is \cite{Brown:1988kg,Kaloper:2022oqv,Kaloper:2022utc}
\be
S_{\tt bounce} 
=  {S}_{out} - {S}_{in} - \pi^2 a^3 {\cal T}_A \, ,
\label{bounce1dact}
\ee
with (where $k \in \{out,in\}$) 
\be
S_k = 18 \pi^2 
\frac{\mpl^4}{\Lambda_{k}} \Bigl( \frac23 -  \zeta_{k} \bigl({1 - \frac{\Lambda_{k} a^2}
{3\mpl^2}}\bigr)^{1/2} + \frac{ \zeta_{k}}{3} \bigl(1 - \frac{\Lambda_{k} a^2}{3\mpl^2}\bigr)^{3/2} \Bigr) \, .
\label{integrals}
\ee
Its value depends on the membrane radius at nucleation $a$, which in turn depends on the microscopic parameters 
and $\Lambda$ according to
\be
\frac{1}{a^2} = \frac{\Lambda_{out}}{3\mps} + \Bigl(\frac{{\cal T}_{\cal A}}{4 \mps}\Bigr)^2 
\Bigr(1 + q \Bigr)^2 
= \frac{\Lambda_{in}}{3\mps} + \Bigl(\frac{{\cal T}_{\cal A}}{4 \mps}\Bigr)^2 
\Bigr(1 - q \Bigr)^2 \, .
\label{radii}
\ee
From this formula we deduce there are two regimes of bubble nucleation for a fixed set of parameters, depending on which term 
on the RHS of (\ref{radii}) the dominant contribution to the membrane radius comes from. The boundary 
between the two regimes is controlled by the critical value of the cosmological constant,
roughly set by $ \Lambda_{*} \simeq 3 (\frac{{\cal T}_{\cal A}}{4 \mpl})^2 (1 + q )^2$. 

To infer a more precise description, we can rewrite the bounce action (\ref{bounce1dact}) in terms of the
{\it out} cosmological constant, membrane charge and tension, and the cutoff scale ${\cal M}_{\tt UV}$. 
First, we can evaluate (\ref{integrals}) by eliminating the square root terms on the RHS using the junction
conditions (\ref{diffroots}). Next we express $\Lambda_{in}$ in terms of $\Lambda_{out}$ and 
membrane charge ${\cal Q}$ 
using (\ref{totalcc}) and the second of (\ref{metricjc}). 
Finally we eliminate powers of the membrane radius at nucleation $a$ using Eq. 
(\ref{radii}).  
Then we can consider specific limits of this action,
e.g. fixing the tension ${\cal T}$ and varying ${\cal Q}$ and $\Lambda_{out}$ relative to it to explore the 
possible tunneling regimes mediated by the instantons of Fig. (\ref{fig1}). 

It is tempting to take a shortcut and merely focus on the leading order terms in this panoply of pieces
in the limits $\Lambda \rightarrow \infty$ and $\Lambda \rightarrow 0$ to get the essential behavior 
of the bounce action (\ref{bounce1dact}) while skipping the algebraic tedium. This actually works 
in the limit $\Lambda \rightarrow 0$. However the limit $\Lambda \rightarrow \infty$ is more delicate. 
The reason is that the $a(\Lambda)$ dependence in (\ref{radii}) and the 
terms $\propto {\cal T}_{\cal A} a$, which the 
bounce action  (\ref{bounce1dact}) depends on after the square roots in  (\ref{integrals}) are evaluated using
 (\ref{diffroots}) show that $\Lambda \rightarrow \infty$ and $\Lambda = 0$ are branch points of the
 bounce action viewed as a function of $\Lambda$ (this can also be seen
 in scalar field tunneling in, e.g. \cite{Parke:1982pm}). In particular, although $a(\Lambda)$ vanishes 
 as $\Lambda \rightarrow \infty$ in (\ref{radii}), the terms $\propto {\cal T}_{\cal A} a$ in the expression for
 the bounce action get multiplied by positive powers of $\Lambda$, and hence may not be negligible. 
Thus it is prudent to determine the exact form of $S_{\tt bounce}$ before taking the limits. 

The calculation is straightforward albeit tedious; a simplifying step is to write the
terms $\frac{ \zeta_{k}}{3} \bigl(1 - \frac{\Lambda_{k} a^2}{3\mpl^2}\bigr)^{3/2} = 
\mp \frac{{\cal T}_{\cal A}}{4\mps}\bigl(1 \pm q \bigr)  \bigl(1 - \frac{\Lambda_{k} a^2}{3\mpl^2}\bigr)$, 
where the upper sign on the RHS corresponds to $k = out$ and the lower sign to $k = in$, respectively, given 
our conventions and definitions here. This gives, after straightforward steps, 
\ba
S_{\tt bounce} &=& 12 \pi^2 \frac{\mpl^4}{\Lambda_{out}} \Bigl(1+\frac{{\cal T}_{\cal A}a}{4\mpl^2}(1+q)\Bigr) 
- 12 \pi^2 \frac{\mpl^4}{\Lambda_{in}} \Bigl(1-\frac{{\cal T}_{\cal A}a}{4\mpl^2}(1-q)\Bigr) \, , \nonumber \\
\frac{{\cal T}_{\cal A}a}{4\mpl^2} &=& \Bigl( \frac{\frac{3}{\Lambda_{out}}(\frac{{\cal T}_{\cal A}}{4\mpl})^2}{1+\frac{3}{\Lambda_{out}}(\frac{{\cal T}_{\cal A}}{4\mpl})^2(1+q)^2} \Bigr)^{1/2} \, , ~~~~~~~~~ 
\Lambda_{in} = \Lambda_{out} + \frac{3{\cal T}_{\cal A}^2}{4\mpl^2} q\, .
\label{bouncepoles}
\ea
Here, of course, $q<0$ since we are focusing on transitions which reduce $\Lambda_{out}$.
It is now clear that $\Lambda_{out} \rightarrow \infty$ and $\Lambda_{out} = 0$ are branch
points. To take the limits it is further convenient to factorize this equation as a product of poles and
functions which only include the branch points. A shortcut is to bring (\ref{bouncepoles}) under
a common denominator, substitute $\Lambda_{out} = \Lambda_{out}(a)$, which turns the 
numerator into a polynomial in $a$, and factorize the polynomial. We obtain
\ba
&& ~~ S_{\tt bounce} =
\frac{18 \pi^2 \mpl^4 {\cal T}_{\cal A}}{\Lambda_{out} \Lambda_{in}  a} \Bigl(1+\frac{{\cal T}_{\cal A}a}{4\mpl^2}(1+q)\Bigr) \Bigl(1-\frac{{\cal T}_{\cal A}a}{4\mpl^2}(1-q)\Bigr) \, , \nonumber \\
&& \frac{{\cal T}_{\cal A}a}{4\mpl^2} = \Bigl( \frac{\frac{3}{\Lambda_{out}}(\frac{{\cal T}_{\cal A}}{4\mpl})^2}{1+\frac{3}{\Lambda_{out}}(\frac{{\cal T}_{\cal A}}{4\mpl})^2(1+q)^2} \Bigr)^{1/2} \, , ~~~~~~ 
\Lambda_{in} = \Lambda_{out} + \frac{3{\cal T}_{\cal A}^2}{4\mpl^2} q\, .
\label{bouncefactors}
\ea
Lastly, we can eliminate $\Lambda_{out}, \Lambda_{\it in}$ using Eq. (\ref{radii}), where we obtain
\ba
\Lambda_{out} &=& \frac{3\mps}{a^2} \Bigl(1-\bigl(\frac{{\cal T}_{\cal A}a}{4\mps}\bigr)(1+q)\Bigr)
\Bigl(1+\bigl(\frac{{\cal T}_{\cal A}a}{4\mps}\bigr)(1+q) \Bigr) \, , \nonumber \\
\Lambda_{in} &=& \frac{3\mps}{a^2} \Bigl(1-\bigl(\frac{{\cal T}_{\cal A}a}{4\mps}\bigr)(1-q)\Bigr) 
\Bigl(1+\big(\frac{{\cal T}_{\cal A}a}{4\mps}\bigr)(1-q) \Bigr) \, ,
\label{lambdas}
\ea
after factorizing the
squares. Substituting these into (\ref{bouncefactors}) we finally find 
\be
S_{\tt bounce} =
\frac{2 \pi^2 a^3 {\cal T}_{\cal A} }{\Bigl(1-\frac{{\cal T}_{\cal A}a}{4\mpl^2}(1+q)\Bigr) \Bigl(1+\frac{{\cal T}_{\cal A}a}{4\mpl^2}(1-q)\Bigr)} \, , ~~~~ 
\frac{{\cal T}_{\cal A}a}{4\mpl^2} = \Bigl( \frac{\frac{3}{\Lambda_{out}}(\frac{{\cal T}_{\cal A}}{4\mpl})^2}{1+\frac{3}{\Lambda_{out}}(\frac{{\cal T}_{\cal A}}{4\mpl})^2(1+q)^2} \Bigr)^{1/2} 
\, .
\label{bouncefactorsa}
\ee

It is now clear that for the $dS \rightarrow dS$ transitions, the bounce action (\ref{bouncefactors}) remains
nonnegative. Given the positivity of the cosmological constants and the tension, and $q<0$, 
the only way it could ever be negative is if the last factor is negative, or alternatively, if
$\frac{{\cal T}_{\cal A}a}{4\mpl^2}(1+q) > 1$. But given the definition of $\frac{{\cal T}_{\cal A}a}{4\mpl^2}$ in  (\ref{bouncefactorsa}), we see that this can't happen: 
\be
\frac{{\cal T}_{\cal A}a}{4\mpl^2}(1+q) = \Bigl( 
\frac{\frac{3}{\Lambda_{out}}(\frac{{\cal T}_{\cal A}}{4\mpl})^2 }{\frac{1}{ (1+q)^2}+\frac{3}{\Lambda_{out}}(\frac{{\cal T}_{\cal A}}{4\mpl})^2}  \Bigr)^{1/2} \le 1 \, .
\label{condeq}
\ee
Using these equations we also see that if the cosmological constant dependent term 
dominates the RHS of (\ref{radii}) -- i.e. in the limit $\Lambda_{out} \rightarrow \infty$ -- 
the membrane's radius at nucleation is $a \simeq \frac{\sqrt{3} \mpl}{\sqrt{\Lambda_{out}}}$ and 
$\frac{{\cal T}_{\cal A}a}{4\mpl^2}(1+q) 
\simeq \frac{{\cal T}_{\cal A}}{4\mpl^2} \frac{\sqrt{3} \mpl}{\sqrt{\Lambda_{out}}}(1+q) < 1$. Thus 
\be
S_{\tt bounce} \simeq   
{2\pi^2 a^3 {\cal T}_{\cal A}} \simeq  \frac{6 \sqrt{3}\pi^2 \mpl^3 {\cal T}_{\cal A}}{ (\Lambda_{out})^{3/2}}  \, . 
\label{limitsb}
\ee
When ${\cal T}_{\cal A} < {\cal M}_{\tt UV}^3$ and $\Lambda_{\tt out} \simeq {\cal M}_{\tt UV}^4 \simeq \mpl^4$,
this gives\footnote{The factor of $6\sqrt{3}\pi^2$ can be easily 
compensated initially by picking the scale $\mu = {\cal T}_{\cal A}^{1/3} < {\cal M}_{\tt UV}/5$. {This may require appropriately reducing charges to keep $|q|<1$; we will return to the precise details in later work.}} 
$S_{\tt bounce} \simeq   
\frac{{\cal T}_{\cal A}}{\mpl^3} < 1$. This is the initial regime which we model-build to be in\footnote{Or to never be in this regime, if we prefer to eventually rely on anthropics, see the discussion later on.}, since in this regime
an initial de Sitter space with a large cosmological constant ``boils" the bubbles of the smaller cosmological constant
that can start populating the landscape. As we noted in the introduction, this is the regime of barrier-less tunneling,
where $dS \rightarrow dS$ decays are unsuppressed. Also, (\ref{limitsb}) grows bigger as $\Lambda_{\tt out}$ 
decreases, which means that de Sitter spaces with the largest $\Lambda$ decay faster than those with
a smaller $\Lambda$. This shows the trend of evolution: fast decay of the large $\Lambda$'s and increased
stability of subsequent lower $\Lambda$ spaces. Clearly, if for any reason the ``boiling" stage is pushed above the cutoff, the theory would not be under control in that regime, and this regime could not
be invoked to set an initial population of $\Lambda$'s. If this were realized, the landscape can turn into a desert. 

This regime ends when the total cosmological constant discharges enough so that the second term in (\ref{radii}) dominates. 
For $|q|<1$, that occurs when $\Lambda  < \Lambda_* = 3(\frac{{\cal T}_{\cal A}}{4 \mpl})^2 $. The discharge proceeds by the 
$|q|<1$ instanton in Fig (\ref{fig1}), for which $(\zeta_{out}, \zeta_{in}) = (-,+)$. The action (\ref{bounce1dact}) gradually asymptotes to 
\be
S_{\tt bounce}  \simeq \frac{24\pi^2 \mpl^4}{\Lambda_{out}} 
\Bigl(1- \frac{8}{3} \frac{\mpl^2  \Lambda_{out}}{ {\cal T}_{\cal A}^2} \Bigr)\, ,
\label{familiars2}
\ee 
in this limit, with $S_{\tt bounce} > 0$ because $\Lambda < 3 \mpl^2  \Bigl(\frac{{\cal T}_{\cal A}}{4 \mpl^2}\Bigr)^2$. This action asymptotes to a pole at $\Lambda_{out} = 0$, and leads to the decay rate of Eq. (\ref{atrrate}), 
$\Gamma \simeq \exp\bigl(-{24 \pi^2 \mpl^4}/{\Lambda_{out}}\bigr)$. Just like the discharges are rapid when 
$\Lambda > 3(\frac{{\cal T}_{\cal A}}{4 \mpl})^2$, they are very slow when $\Lambda  < 3(\frac{{\cal T}_{\cal A}}{4 \mpl})^2$. 
As a result, {the} total evolution which results from combining the stages controlled 
by (\ref{limitsb}) and (\ref{familiars2}) , if both
can be fit below the cutoff in the same effective theory, would favor {the} smallest values of $\Lambda_{\tt terminal}$. The ``boiling" stage {is} setting up the distribution and {the} ``braking" stage preserving it, 
and therefore {in tandem they} provide a framework for naturally solving the cosmological constant problem. On the other hand, if the
``boiling" regime were completely excised, being pushed above the cutoff, the resulting 
landscape could be very desolate with naturally large initial $\Lambda$. In this case, the decay of the 
cosmological constant may be very slow, and if the membrane charges are small, it would need to go
through many steps until reaching the terminal $\Lambda$ values near zero. At practical times, the 
distribution of  $\Lambda$ would be biased toward larger values, and the prospect of ultimately empty universe
\cite{Abbott:1984qf} would loom large. 

In the other case, when $|q|>1$, the ``braking" regime starts when $\Lambda <\frac{\mpl^6 {\cal Q}^2_{\cal A}}{12 {\cal T}_{\cal A}^2}$, or when $\Lambda <\frac{4\mpl^2 \lambda^2 {\cal Q}^2_{\cal A}}{3 {\cal T}_{\cal A}^2}$, depending on whether
linear or quadratic terms dominate $q$. The
discharges are mediated by the $|q|>1$ instanton in Fig (\ref{fig1}), for which $(\zeta_{out}, \zeta_{in}) = (+,+)$. For this instanton, 
in the bounce action (\ref{bounce1dact}), (\ref{integrals}), the leading terms in (\ref{integrals}) cancel out completely for both ``in" and
``out" contributions, and the sub-leading terms converge to (see, e.g. 
\cite{Coleman:1980aw,Brown:1987dd,Brown:1988kg,Kaloper:2022oqv,Kaloper:2022utc}) 
\be
S_{\tt bounce} \simeq \frac{27\pi^2}{2} \frac{{\cal T}_{\cal A}^4}{(\Delta \Lambda)^3}\, .
\label{familiarso}
\ee 
In this regime the decay rate saturates for a broad range of $\Lambda$ as the cosmological constant decreases. When the quadratic flux terms dominate in $q$, 
the relative stability of de Sitter spaces with small cosmological constant 
is set by the ratio\footnote{The appearance of $\Lambda_{\tt QFT}$ follows from requiring natural screening of 
vacuum energy by fluxes \cite{Bousso:2000xa,Kaloper:2023kua}. If the linear flux terms dominate, 
we'd find ${\cal T}_{\cal A}^4 > \mpl^{6} {\cal Q}_{\cal A}^{3}$ instead of (\ref{stabbound}). } 
${\cal T}_{\cal A}^4/(\Delta \Lambda)^3 \simeq {\cal T}_{\cal A}^4/(2 \Lambda_{\tt QFT})^{3/2} {\cal Q}_{\cal A}^{3}$, 
with the decay rate approaching (\ref{constrate}). For a natural value of the screened initial vacuum energy 
$\Lambda_{\tt QFT} \simeq {\cal M}_{\tt UV}^4$, this immediately shows that we need, somewhat loosely,
\be
{\cal T}_{\cal A}^4 > (2 \Lambda_{\tt QFT})^{3/2} {\cal Q}_{\cal A}^{3} \, , 
\label{stabbound}
\ee
to have a chance for sufficient longevity of de Sitter regions with small cosmological constant, necessary to fit a realistic late universe cosmology.
If the tension were too low, the small curvature de Sitter spaces could decay too fast. However, since the rate is approximately constant, when
(\ref{stabbound}) holds, and if the ``boiling" regime (\ref{limitsb}) is not too long, 
the discharges can produce a multiverse with 
all values of $\Lambda_{\tt terminal}$ approximately equally likely, and long lived. 
If this happens, then invoking anthropics can be used to address the observed smallness of the cosmological constant. 
As we will see below, this can naturally occur when all flux discharge processes are dominated by quadratic or higher order flux terms. 

\section{WGC versus Discharges}

We now impose the WGC bounds of Sec. 2 on the discharge dynamics of the previous section. We will
normalize the inequalities (\ref{wgcmeme}), (\ref{wgcmemm}) using the Planck scale instead of Newton's constant,
ignoring the numerical factor of $\sqrt{8\pi} \simeq 5$, thus working with the original normalizations introduced in \cite{Arkani-Hamed:2006emk}.
The ${\cal O}(1)$ numerical factors will be of little consequence in this work, although in general one should be
careful with their accounting since they can affect normalization of some physical parameters, as for example the
overall normalization of the bounce action, the 
duration of slow roll inflation and so on \cite{DAmico:2017cda,DAmico:2018mnx}. In any case, the electric and magnetic weak
gravity bounds which we will use are 
\be
\mpl \frac{{\cal Q}}{{\cal T}} \ge 1 \, ,
\label{wgcmemep}
\ee
and
\be
 {\cal Q} \mpl  \ge {\cal M}_{\tt UV}^3 \, .
\label{wgcmemmp}
\ee

In addition, following the approach of \cite{Kaloper:2023kua}, we will impose a bound on the flux variation for each
type of flux involved in screening and discharge. The reason for this is that in hindsight, when the $4$-forms are 
generalized by adding a dynamical longitudinal mode and a mass term, which arises naturally whenever the $4$-forms
realize monodromy field theories in $4D$, as in 
\cite{Silverstein:2008sg,McAllister:2008hb,Kaloper:2008qs,Kaloper:2008fb,Dong:2010in,Kaloper:2011jz,Kaloper:2014zba,Marchesano:2014mla,McAllister:2014mpa,Kaloper:2016fbr,Montero:2017yja,Buratti:2018xjt}, in the axial gauge the longitudinal modes are monodromy-spanning ``axions", whose
total range must be limited by at least the requirement that their energy density does not exceed the Planckian energy, so the effective
theory with gravity remains under control. Depending on the specifics of the theory, the bounds could be even tighter. Here, imagining that the
effective theory enjoys protection from the gauge symmetries of the $4$-form sectors, both continuous and discrete, we will require that it remains below the cutoff scale 
\be
|| T^{\mu}{}_{\nu}(4-{\rm form}) || \la {\cal M}_{\tt UV}^4 \la \mpl^4 \, ,
\label{cutoffbound}
\ee
where by $|| T^{\mu}{}_{\nu}(4-{\rm form}) ||$ we mean the operatorial norm of the stress energy tensor of the $4$-form 
sector, i.e. its largest eigenvalue. 
With this in place, we are ready to find the implications of these bounds on the dynamics of screening and discharge. Conveniently,
the technical aspects of this analysis are simplified by separately considering the purely quadratic flux case, as an avatar of frameworks
where the linear flux term is sub-leading, and purely linear term, without any loss of generality.

\subsection{Quadratic Flux Dominance}

We will explicitly work with a single species of membranes for the most part, since the nucleations proceed one bubble at a time.
However we bear in mind that, to be able to approach the observably allowed values of the terminal cosmological constant, 
we need a multiplicity of 
different membranes once we impose the field theory cutoff on the flux range \cite{Bousso:2000xa,Kaloper:2023kua}. This means that in the 
formulas below we should really replace expressions like 
$N^2_{\cal A} {\cal Q}^2_{\cal A}$ by  $\sum_i N^2_{i} {\cal Q}^2_{i}$ etc. This in turn 
means, that in our comparison of the 
cosmological constant to be cancelled and the cutoff, there is an in-principle multiplicative species factor, counting
each flux that contributes to $\Lambda_{\tt total}$. Since it is $\la {\cal O}(100)$ we will ignore it in what follows.
\begin{figure}[thb]
    \centering
    \includegraphics[width=11cm]{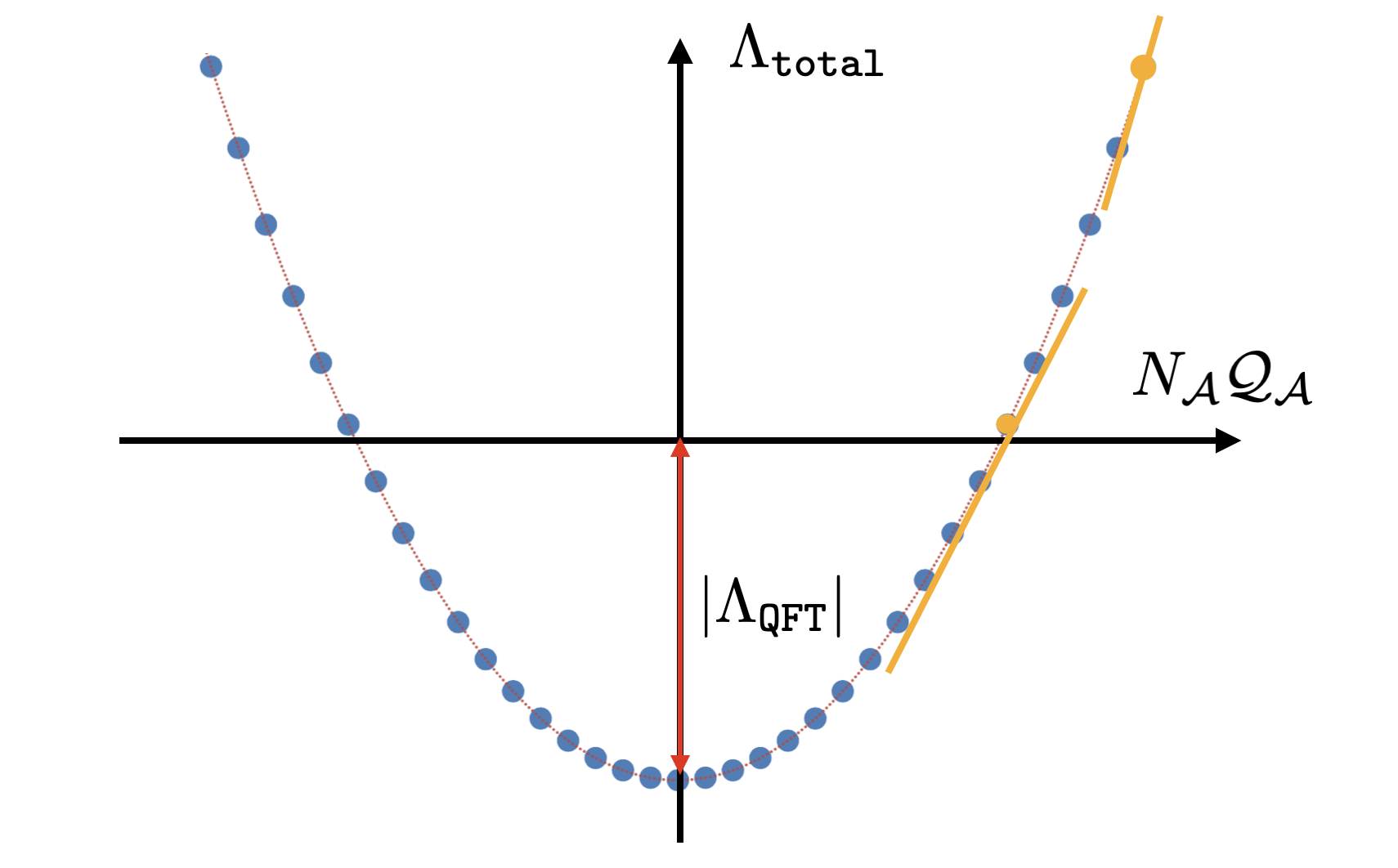} 
     \caption{$\Lambda$-parabola, depicting the spectrum of $\Lambda$ as a function of the screening flux. In the full multidimensional 
     flux space, $\Lambda$ is a paraboloid, and here we depict its projection to a single coordinate plane. The gold lines are
     tangents to the parabola whose slope is $q$, which controls the discharge process. The discrete points are the actual values of the quantized fluxes and the corresponding cosmological constant of the $\Lambda$-discretuum.}
    \label{fig2}
\end{figure}
When the screening terms in $\Lambda_{\tt total}$ are dominated by the quadratic flux contributions, such that
\be
\Lambda_{\tt total} = \Lambda_{\tt QFT} + 2 \sum \lambda_i^2 + \ldots \simeq -| \Lambda_{\tt QFT}| + 2 \sum_i \lambda_i^2  \, ,  
\label{totalccq}
\ee
we must take $\Lambda_{\tt QFT} < 0$ to have a chance to cancel it \cite{Brown:1987dd,Brown:1988kg,Bousso:2000xa}. Then, 
for a natural value of $\Lambda_{\tt QFT} \sim {\cal M}_{\tt UV}^4 \sim \mpl^4$, the dynamics of discharge produces a nested system of bubbles
bounded by membranes. Nucleation processes are controlled by Eqs. (\ref{diffroots}) - (\ref{radii}), and, crucially, by the value of 
$q$. In the limit when quadratic terms dominate, $q$ is given by 
\be
q_i = \frac83 \frac{\mpl^2 \lambda_i {\cal Q}_i}{{\cal T}^2_i} \, ,
\ee
for each individual flux $\lambda_i$. The parameter $q_i$ is proportional to the slope of the tangent to the ``spectral parabola" as depicted 
in Fig. (\ref{fig2}). 

Since fluxes are quantized, $\lambda_i = \frac12 N_i {\cal Q}_i$ ($1/2$ comes from our normalization of $\lambda$). Then, plugging this 
into the formula for $q_i$ yields $q_i = \frac43 N_i \frac{\mpl^2 {\cal Q}_i^2}{{\cal T}_i^2}$, or, using 
$\gamma_{\tt WGC} = \mpl {\cal Q}_i/{\cal T}_i$,
\be
q_i = \frac43 N_i \gamma_{\tt WGC}^2 \, ,
\label{qflux}
\ee
where $\gamma_{\tt WGC}$ is precisely the ratio of charge to tension in Planck units, which is subject to 
the electric weak gravity bound (\ref{wgcmemep}). Thus, if WGC is obeyed by a membrane $``i"$, 
$\gamma_{\tt WGC} > 1$, and since we must screen a natural vacuum energy $\Lambda_{\tt QFT}$ with multiple units of flux, $N_i > 1$. Therefore
$q > 1$ for any type of membrane obeying WGC, for all transitions which occur until $\Lambda$ reaches its smallest positive value. 
As a result, the discharge processes of the natural vacuum energy by emission of these
membranes can only proceed by the instanton with $|q|>1$ of Fig. (\ref{fig1}). This means, the bounce action for these processes generically asymptotically 
approaches Eq. (\ref{familiarso}) as the cosmological constant diminishes, which remains
a good approximation for much of the discharge sequence. 

The relevant inequalities to check further are (\ref{stabbound}) and (\ref{wgcmemmp}). In fact, we can rewrite all three of these inequalities in terms
of dimensionless ratios, as follows (for $\Lambda_{\tt QFT} \sim {\cal M}_{\tt UV}^4$):
\ba
\Bigl(\frac{{\cal T}_{i}}{\mpl^3} \Bigr)^4 &>& \Bigl(\frac{{\cal M}_{\tt UV}}{\mpl} \Bigr)^6 \Bigl(\frac{{\cal Q}_{i}}{\mpl^2} \Bigr)^{3} \, , ~~~~~~~~ {\rm stability} \, ; \nonumber \\
\Bigl(\frac{{\cal Q}_{i}}{\mpl^2} \Bigr) \, &\ge& \Bigl(\frac{{\cal T}_{i}}{\mpl^3} \Bigr) \, ,  
~~~~~~~~~~~~~~~~~~~~~ {\rm electric ~WGC} \, ; \nonumber \\
\Bigl(\frac{{\cal Q}_{i}}{\mpl^2} \Bigr) \, &\ge& \Bigl(\frac{{\cal M}_{\tt UV}}{\mpl} \Bigr)^{3} \, ,  
~~~~~~~~~~~~~~~~~~~ {\rm magnetic~WGC} \, .
\label{allbounds}
\ea 

Furthermore, since the membrane charge and tension are distributed quantities, we should require that they are 
below the cutoff scale, ${\cal Q}_i < {\cal M}_{\tt UV}^2$ and ${\cal T}_i < {\cal M}_{\tt UV}^3$, so that
they can be reliably included in the sub-cutoff effective description based on the low energy actions which we deploy 
here. In fact, these bounds are redundant: the electric WGC bound in (\ref{allbounds}) follows from the magnetic WGC bound and ${\cal T}_i < {\cal M}_{\tt UV}^3$. However we will retain the electric WGC bound for convenience with calculations below. We note that 
all of these inequalities can be satisfied simultaneously for some ${\cal M}_{\tt UV} \la \mpl$. 
On the other hand, model building `economics' suggests that ${\cal M}_{\tt UV}$ is to be looked for
not too far below $\mpl$ in order to be able to use as few fluxes as possible \cite{Bousso:2000xa}. Add
to this the argument of the previous section about the existence of the ``boiling" stage, which further reaffirms
this expectation. 

Finally, we note that, for as long as $|q|>1$, the 
scale that separates the ``boiling" from the ``braking" stage for quadratic flux is given by
\be
\Lambda_* \simeq \frac43 \frac{\lambda^2 \mpl^2 {\cal Q}^2_{i}}{{\cal T}_{i}^2} \simeq  \frac23 \gamma_{\tt WGC}^2 
{\cal M}_{\tt UV}^4 \, .
\label{boilout}
\ee
When $\gamma_{\tt WGC} > 1$, the critical value $\Lambda_*$ is above the cutoff, and 
``boiling" stage is effectively excised out of the 
quadratic flux effective theory, and so essentially for values of $\Lambda \la {\cal M}_{\tt UV}^4 \sim \mpl^4$
the discharges occur near the edge or during the ``braking" stage (with the decay rate reduction being progressively more efficient as $\Lambda$ decreases), with the bounce action asymptotically approaching the formula 
given in Eq. (\ref{familiarso}), $S_{\tt bounce} \simeq \frac{27\pi^2}{2} \frac{{\cal T}_{i}^4}{(\Delta \Lambda)^3}$.
Since $\Delta \Lambda = 2\lambda \Delta \lambda$, and initially $\lambda \simeq \sqrt{\Lambda} \ga 
\sqrt{\Lambda_{\tt QFT}}$, the initial bounce action will start smaller than the asymptotic value 
$S_{\tt bounce} \simeq \frac{27\pi^2}{2} \frac{{\cal T}_i^4}{(2 \Lambda_{\tt QFT})^{3/2} {\cal Q}_i^3}$, which 
may permit discharges at an approximately uniform rate, independent 
of initial and final cosmological constant
values. The discharges will cease once the flux becomes small enough to obey the stability bound (\ref{stabbound}).
Thus the theory where flux contributions to the cosmological constant are dominated by quadratic terms
may provide an arena where invoking the anthropic principle can be used to address the observed
smallness of the cosmological constant, as in \cite{Bousso:2000xa}.

Higher powers of flux do not affect this conclusion much. If e.g. $2\lambda^2$ is replaced by
${\cal L}(\lambda) = 2 \lambda^2 \bigl(1+ c_3 {\lambda}/{{\cal M}_{\tt UV}^2} + \ldots \bigr)$, as in
the examples of \cite{Kaloper:2023xfl},
and higher order terms are suppressed by the cutoff, or by $\mpl$, these terms will be sub-leading
in the effective theory. If, on the other hand, the suppression is weaker for any single one of them, that term might
compete with the quadratic flux term at large flux, and perhaps even produce novel regimes with tiny 
total cosmological constant, rearranging the effective theory near them but behaving similarly to when the
quadratic dominates \cite{Kaloper:2023xfl}. 

Thus the bottomline is that these processes can discharge the cosmological constant at a nearly 
constant rate from
one value to another before the discharges stop completely, setting essentially an approximately uniform 
distribution of values at late times, without automatically favoring any particular value of $\Lambda$, including
the smallest ones. This sets the stage for invoking anthropic principle. 

\begin{figure}[thb]
    \centering
    \includegraphics[width=11cm]{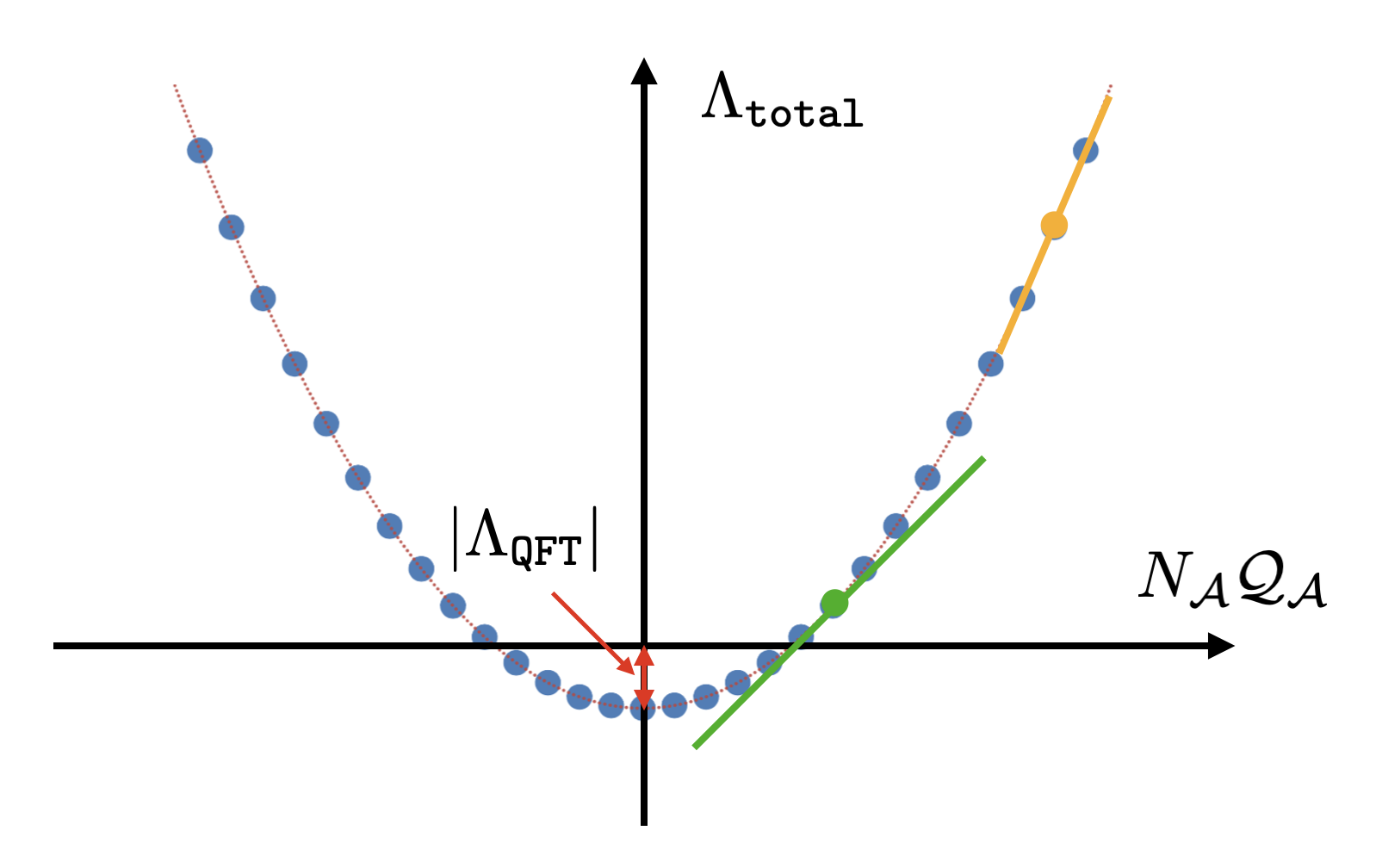} 
     \caption{$\Lambda$-parabola, depicting the spectrum of $\Lambda$ as a function of the screening flux, but for
     smaller values of ${\cal Q}_i$ than in Fig. (\ref{fig2}). As a result, the slope of the tangent as measured by q 
     becomes smaller, and if $|q|<1$ discharges would be mediated by the $|q|<1$ instanton of Fig. (\ref{fig1}). We show
     in the text that this is unfounded when quadratic flux terms dominate.}
    \label{fig3}
\end{figure}

In contrast, in \cite{Kaloper:2022oqv,Kaloper:2022utc,Kaloper:2022jpv,Kaloper:2023xfl,Kaloper:2023kua}
we have been pursuing a framework where the dominant flux terms are linear, and the instantons
which discharge the cosmological constant have $|q|<1$, so that their bounce action asymptotes to a pole
at a tiny value of $\Lambda$, which, as we argued, can favor the smallest $\Lambda$ without 
anthropics. As we noted in the introduction, one might try to adopt similar processes to the
cases when higher powers of the flux dominate, and the linear term is absent. One might think that by arbitrarily 
reducing the membrane charge, this might make $q$ smaller than unity, so that 
the $|q|<1$ instantons of Fig. (\ref{fig1}) take over the discharges. 
If this had been possible, the spectrum of $\Lambda$ as a function of
the fluxes would have been altered, looking like Fig. (\ref{fig3}), where for small 
parent $\Lambda$ the slope of the tangent to the parabola would have been below unity. However, there
are problems with this approach.

Seeing the problem is straightforward. To get $|q|<1$, so that the corresponding instantons are activated, formula
(\ref{qflux}) shows that we must violate the electric weak gravity bound {\it considerably}: solving (\ref{qflux}) for
$N_i$,
\be
N_i = \frac34 \frac{q_i}{\gamma_{\tt WGC}^2} \, , 
\ee
and so if $|q_i| <1$ and $\gamma_{\tt WGC} >1$, we find $N_i <0.75$ - which completely excludes the possibility
of screening {\it any} value of field theory vacuum energy by quantized fluxes. Indeed, to have a chance to screen a natural field theory vacuum energy $\lambda_{\tt QFT} \simeq {\cal M}_{\tt UV}^4$, and then relax the total by subsequent membrane nucleations, 
we need multiple units of flux: $N_i \gg 1$. Hence we need a serious violation of 
the electric weak gravity bound: $|q_i| <1$ implies 
\be
\gamma_{\tt WGC}  < \sqrt{\frac{3}{4 N_i}} \, . 
\label{gammawgc}
\ee
and so if $N_i \gg 1$, 
we find $\gamma_{\tt WGC} \ll 1$. Next, we can ignore the stability bound of Eqs. (\ref{allbounds}), since the bounce action in the ``braking" stage for this case isn't (\ref{familiarso}), but (\ref{familiars2}), and so the 
stability is enforced by the $\Lambda \rightarrow 0$ pole. 

However, if we rewrite $\Lambda_*$ for the case $|q|<1$ as 
\be
\Lambda_* = \frac{3}{16} \frac{{\cal T}_i^2}{\mpl^2} = \frac{3}{16} \frac{{\cal Q}_i^2}{\gamma_{\tt WGC}^2} \, ,
\label{lstar}
\ee
we see that $\gamma_{\tt WGC} \ll 1$ implies that charges are very small: ${\cal Q}_i \simeq \frac{4}{\sqrt{3}} 
\gamma_{\tt WGC} \sqrt{\Lambda_*} \ll \frac{4}{\sqrt{3}}\sqrt{\Lambda_*} $. From Eqs. (\ref{totalcc}) and the second of (\ref{metricjc}) it then follows
that
\be
\Delta \Lambda \simeq 2 \lambda \Delta \lambda = \lambda {\cal Q}_{i} \simeq \frac{4}{\sqrt{3}} 
\gamma_{\tt WGC}  \sqrt{\Lambda_{\tt QFT}} \sqrt{\Lambda_*} \ll
\frac{4}{\sqrt{3}} {\cal M}_{\tt UV}^2 \sqrt{\Lambda_*} 
\simeq \frac{{\cal M}_{\tt UV}^2}{\mpl} {\cal T}_{i} \, .
\label{dellsmall}
\ee
Thus since ${\cal M}_{\tt UV} \sim \mpl$ and ${\cal T}_{i} < {\cal M}_{\tt UV}^3$, in this regime
the individual discharges change the cosmological constant by a tiny amount,
$\Delta \Lambda/\Lambda \ll {\cal T}_{i}/{\cal M}_{\tt UV}^3$. Thus to relax it to nearly zero, 
we need many subsequent transitions during the ``braking" stage, which will be ever more slow due to the 
attractor behavior of (\ref{familiars2}). Such a slow discharge sequence practically stabilizes the
de Sitter background, and it could bring back the specter of the empty universe
problem \cite{Abbott:1984qf}, since many small slow steps could result in difficult reheating\footnote{Dynamics of  
stable bubbles and domain walls with small tensions and charges can also 
be constrained by cosmology \cite{Zeldovich:1974uw}, although
such bounds are not very practical in an empty universe.}. 

The empty universe problem could be averted by adding a single membrane with tension and charge 
which satisfy WGC bounds. This would maintain the option of UV-completing the theory, and
it would mediate faster nucleations in the ``braking" stage, by using the $|q|>1$ nucleation processes 
with the bounce action (\ref{familiarso}). Even if those satisfy the stability bound (\ref{stabbound}), 
the transitions would be generically faster than the ones mediating (\ref{dellsmall}). Discharges mediated
by such a membrane (or membranes) can overtake the
processes which have the attractor behavior and avert the empty universe. However, since the decay rate in
this case may be uniform, this channel could usher the anthropics back. This is the obstacle to some
of the proposals in \cite{jajare}, in using $|q|<1$ instantons when quadratic (and higher power) fluxes dominate. 
The point is not that those small values of $\Lambda$ aren't populated, but {\it how} they are populated. 
Indeed, on general grounds the whole landscape will be populated \cite{Brown:2011ry}, but the details of the 
landscape `demographics' must be looked at on a case to case basis. 

\subsection{Linear Flux Dominance}

\begin{figure}[thb]
    \centering
    \includegraphics[width=8.5cm]{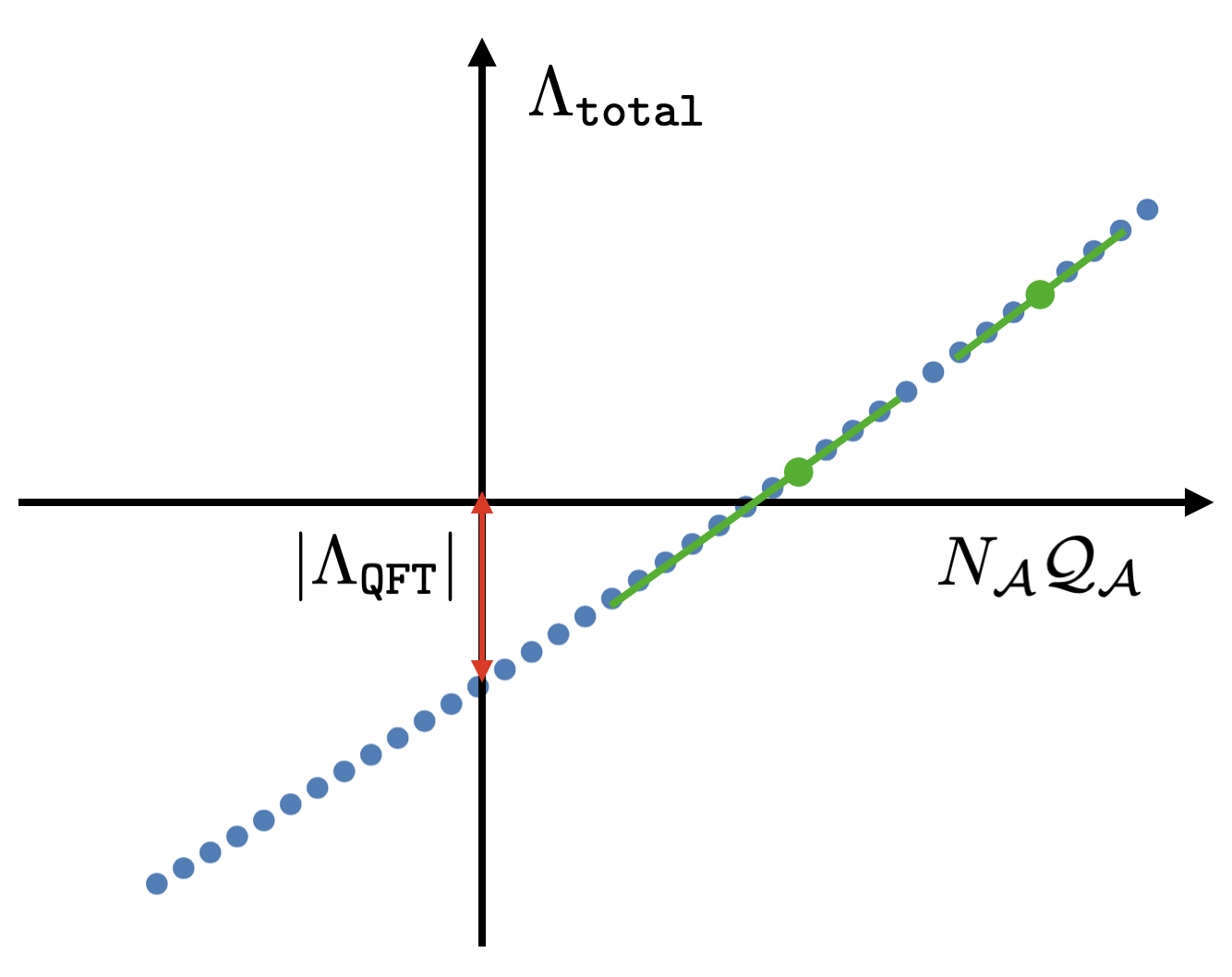} 
     \caption{Almost-linear spectrum of $\Lambda$ as a function of the screening flux, 
     projected onto a single coordinate plane. The slope of the tangent  
     is practically a constant, and when $|q|<1$ the membrane discharges are mediated by the $|q|<1$ instantons
     of Fig. (\ref{fig1}).}
    \label{fig4}
\end{figure}

We have already extensively discussed aspects of $\Lambda$ discharge when linear flux terms dominate in
\cite{Kaloper:2022oqv,Kaloper:2022utc,Kaloper:2022jpv,Kaloper:2023xfl,Kaloper:2023kua}. Here we will revisit 
some of those results with particular attention given to the role of WGC bounds. First off, the screened total
cosmological constant, as e.g. examined in \cite{Kaloper:2023kua}, is
\be
\Lambda_{\tt total} = \Lambda_{\tt QFT} - c_1 {\cal M}_{\tt UV}^2 \lambda + 2 \lambda^2 = 
 \Lambda_{\tt QFT} - \frac{c_1^2}{2} {\cal M}_{\tt UV}^4 + 2 \bigl(\lambda - \frac{c_1}{4}{\cal M}_{\tt UV}^2 \bigr)^2  \, , 
\label{totallincc}
\ee
where we implicitly take the linear term to dominate over the quadratic one, and complete the squares in the
second line for the sake of convenience. In the case of multiple fluxes we can rewrite this as
\be
\Lambda_{\tt eff} =  2 \sum_i \bigl(\lambda_i - \frac{c_{1\,i}}{4} {\cal M}_{\tt UV}^2 \bigr)^2  \, , ~~~~~~~~~~~~~
\Lambda_{\tt eff}  = \Lambda_{\tt total} - \Lambda_{\tt QFT} + \sum_i \frac{c_{1\, i}^2}{2} {\cal M}_{\tt UV}^4 \, .  
\label{efftotallincc}
\ee
The spectrum of values of $\Lambda$ is depicted in Fig. (\ref{fig4}).  
Note that in this case it does not matter if $\Lambda_{\tt QFT}$ is positive or negative (it had to be negative when
quadratic fluxes dominate for screening to work). Due to the fact that linear fluxes dominate, they can screen
$\Lambda_{\tt QFT}$ of either sign. 

In this limit, the slope parameter is
\be
q_i = \frac{2c_{1\,i}}{3} \frac{\mpl^2{\cal M}_{\tt UV}^2{\cal Q}_{i}}{{\cal T}^2_{i}} 
=   \frac{2c_{1\,i}}{3} \frac{\mpl {\cal M}_{\tt UV}^2}{ {\cal T}_{i}} \gamma_{\tt WGC} =  \frac{2c_{1\,i}}{3} \frac{{\cal M}_{\tt UV}^2}{ {\cal Q}_{i}} \gamma_{\tt WGC}^2 \, .
\label{slope}
\ee
The equation for $\Lambda_*$ is still given by the expression (\ref{lstar}), $\Lambda_* = \frac{3}{16} {{\cal T}_i^2}/{\mpl^2} = \frac{3}{16} {{\cal Q}_i^2}/{\gamma_{\tt WGC}^2}$. 
The important point is that now the parameter $q_i$ is completely independent of the units of flux $N_i$ - which can be as large as one wishes while $q_i$ is still fixed. 
We can keep the charge near the cutoff, 
${\cal Q}_i \sim {\cal M}_{\tt UV}^2$, and $\gamma_{\tt WGC} \la 1$, but close to the bound, and ensure that
$|q_i| < 1$ by choosing $c_{1~i} < 1$. When $\gamma_{\tt WGC} \la1$, $\Lambda_*$ might thus be mere few 
units of charge above zero, but faster discharges during the ``boiling" regime could discharge it to near zero. 

Then, to satisfy WGC, we in principle need to have only one
charge per gauge group which satisfies the electric bound. Saturating the bound is acceptable, when the 
charge is light enough. So for each gauge group we need one of the membranes to obey 
$\gamma_{\tt WGC} \simeq 1$. From Eq. (\ref{slope}), for membranes with $\gamma_{\tt WGC} \simeq 1$
we need $2 c_{1~i} {\cal M}_{\tt UV}^2 < 3 {\cal Q}_i$ to get $|q_i| < 1$.
This together with ${\cal Q}_{i} \le {\cal M}_{\tt UV}^2$
constrains the charge to $2 c_{1~i} {\cal M}_{\tt UV}^2/3  < {\cal Q}_i \le {\cal M}_{\tt UV}^2$, which can be met
by $c_{1~i} \le 1$. Meeting these bounds may be easiest done near the Planckian
cutoff, $c_{1\,i} \mpl \sim {\cal M}_{UV}$, which can also keep ${\cal T}_{i} < {\cal M}_{\tt UV}^3$ 
and so retain an episode of ``boiling" in the theory below the cutoff. So a membrane
with a charge of order of the cutoff and tension somewhat smaller 
could also yield $|q_i| \la 1$ while marginally satisfying WGC bounds. For other membranes which carry the gauge 
charge of the same group, we can then violate the electric weak gravity bound, that will make
achieving $|q_i| < 1$ much easier. This can happen if, for example,
those membranes are domain walls separating multiple vacua after a symmetry breaking at low energies,
which could be viewed analogously to particles with fractional charges in QFT. 

Alternatively, we may even allow a membrane whose charge to tension ratio obeys the electric weak gravity
bound to participate in the discharge of $\Lambda$ even if it has $|q_i| \ga 1$, as long as there are 
other species of membranes with $|q_i| < 1$ (which may violate the WGC bounds). The reason is that although
the processes for this one specific discharge channel are mediated by the $|q| \ga 1$ instanton, which uniformizes 
the distribution of the $\Lambda$ values which are linked by these transitions, there is many more channels which
proceed via the $|q|<1$ instantons. Those can still bias the overall distributions of $\Lambda$  
towards the smallest possible values, and have larger charges that require fewer steps to get the cosmological
constant close to zero. 
Once near zero, those small values remain extremely stable. If a $|q| \ga 1$ channel is
present, those values might not be absolutely stable: they could decay, for example to regions $\Lambda < 0$ 
{\it eventually}. But as we have seen, once a region of the universe ends up in the ``braking" regime of either
$|q|<1$ or $|q| \ga 1$ instanton discharge, it is very stable and very long lived. Yet, when the terminal distribution
of $\Lambda$ is biased towards smallest possible values, we may still avoid invoking the anthropic arguments
to explain why the cosmological constant is not huge. Exactly how close to zero it can be is controlled by the
model building aspects of the theory, which we described in  some detail in 
\cite{Kaloper:2022oqv,Kaloper:2022utc,Kaloper:2022jpv,Kaloper:2023xfl,Kaloper:2023kua}. 
We direct an interested reader to those references for additional information. 

\section{Summary}

In this paper we have examined in detail implications of the weak gravity conjecture for the mechanisms for discharging cosmological constant via membrane nucleations. This is a natural and interesting question, given the role which
the WGC bounds play in blocking the existence of eternal event horizons in gravity theories in order to protect unitarity. 
In the frameworks where the
cosmological constant is screened by $4$-form fluxes, and then the total background value of $\Lambda$ is  
discharged away by the nucleation of membranes, stable eternal de Sitter spaces do not even exist. In fact, starting with a theory 
which has fluxes and membranes, the only way to recover an eternal de Sitter is to decouple all of the membranes
in the theory by, e.g., sending their tensions to infinity. But this would violate the WGC bounds 
completely; compliance with the conjecture rules out eternal de Sitter just like the WGC bounds 
applied to charged particles rules out
eternal extremal black holes \cite{Arkani-Hamed:2006emk}. 
Conversely, in the example where quadratic flux terms control the discharge processes 
we saw that if the WGC bounds are violated, de Sitter space will not reach near-Minkowski limit
unless the theory is fine tuned. From this point of view, an eternal de Sitter geometry
is really analogous to a remnant, with regions forever removed from a dweller in the space. 

The details of the WGC bounds, when combined with naturalness of the initial, maximal cosmological constant,
place limits on the decay processes of de Sitter space and the cosmological constant which sources it. The possible 
outcomes fall in two different classes. When the flux terms which control the screening and discharge 
of the cosmological constant are dominated by quadratic and higher order terms, 
the bounds from weak gravity conjecture and naturalness point toward anthropic scenarios. 
Interestingly, even if the WGC bounds are deliberately violated, the discharge rates still do not easily 
pick a small $\Lambda$. On the other hand, if the theory involves linear flux terms, which dominate 
below the cutoff, anthropics could be avoided, because a large cosmological constant naturally decays toward smaller values, with non-uniform decay rates. 

This is because the evolution is comprised of two discharge regimes: the ``boiling" stage, followed by a 
``braking" stage. For the cases when the linear fluxes are present and dominant, with WGC-compliant branes, the 
narrower ``boiling" stage processes have fast discharge rates, which generate the descendant regions with curvatures biased towards the smallest possible values of $\Lambda>0$. The 
subsequent ``braking" stage in turn slows down the most the decay rates of regions with smallest positive $\Lambda$ after ``boiling" has ended. Together, these stages produce a distribution of $\Lambda$ which 
is biased towards the smallest values. 

Conversely, if the controlling fluxes are dominated by quadratic or higher order terms, purposefully violating the WGC bounds may reproduce the enhanced ``braking" at small values of $\Lambda$, making de Sitter patches 
with small $\Lambda$ more stable than the strongly curved ones. 
However, the price to pay is that charges must be very small if the screening and discharge adjustment are
to be natural. This affects the ``boiling" stage, which could 
resurrect the empty universe problem. 
Further, by current lore, UV-completing the theory needs the means to maintain WGC. 
This can be done by adding a membrane which satisfies the electric WGC bound, charged under the same gauge 
group as the membranes which are adjusting $\Lambda$, one per gauge group. If gravity is universal, these
membranes should also partake in the 
cosmological constant adjustment. If they obey WGC bounds, their vacuum energy would be 
dominated by quadratic flux terms and so they would yield discharge processes which would be
uniform, since it would be mediated by the $(+,+)$ instanton. These processes can be faster than
the discharges using WGC-violating membranes, and flatten the distribution of 
terminal $\Lambda$ at the small $\Lambda$ end. This would usher back the anthropics. 

Note that our investigation of the discharges was consistently carried out below the cutoff of the effective theory,
which avoids the direct confrontation with quandary that is the wormhole regime 
\cite{Coleman:1988tj,Fischler:1988ia,Klebanov:1988eh,Coleman:1989ky}. In this sense, the WGC limits are
useful, since they ``regulate" the boundary conditions which quantum gravity imposes on the 
low energy effective theory, with some confidence that the phenomena retained because 
they obey the WGC bounds are meaningfully accounted for.

Regarding the final numerical values of $\Lambda$, precisely how small those 
can be depends on the model building details, as noted in previous work
\cite{Kaloper:2022oqv,Kaloper:2022utc,Kaloper:2022jpv,Kaloper:2023xfl,Kaloper:2023kua}. 
More precise model building is required to give a more specific answer. 
Finally, note that even if the various regimes occur concurrently, i.e. if the discharge processes are 
more diversified, involving both processes mediated by
$|q|>1$ and $|q|<1$ instantons, as long as at least some channels are dominated by linear fluxes the 
spectrum of $\Lambda$ will be skewed toward the smallest values. A more detailed investigation
to explore phenomenologically viable scenarios is therefore warranted. 

\vskip1cm

{\bf Acknowledgments}: 
We thank G. D'Amico for useful discussions. We especially thank F. Pedro for questions and comments
which helped sharpen the arguments here. 
NK is supported in part by the DOE Grant DE-SC0009999. AW is partially supported by the Deutsche Forschungsgemeinschaft under Germany's 
Excellence Strategy - EXC 2121 "Quantum Universe" - 390833306.

\end{document}